\documentclass[draft]{agujournal2019}
\usepackage{url} %this package should fix any errors with URLs in refs.
\usepackage{lineno}
\usepackage[inline]{trackchanges} %for better track changes. finalnew option will compile document with changes incorporated.
\usepackage{soul}
\usepackage{amsfonts,amsmath,amssymb,amsthm}
\draftfalse

\journalname{JGR: Space Physics}

\newcommand{\ssr}{    {Space Sci. Rev.}}

\begin{document}

%%%%%%%%%%%%%%%%%%%%%%%%%%%%%%%%%%%%%%%%%%%%%%%
%  TITLE
%
% (A title should be specific, informative, and brief. Use
% abbreviations only if they are defined in the abstract. Titles that
% start with general keywords then specific terms are optimized in
% searches)
%
%%%%%%%%%%%%%%%%%%%%%%%%%%%%%%%%%%%%%%%%%%%%%%%

% Example: \title{This is a test title}

\title{Inferring lunar wake potentials from electron phase space densities}

%%%%%%%%%%%%%%%%%%%%%%%%%%%%%%%%%%%%%%%%%%%%%%%
%
%  AUTHORS AND AFFILIATIONS
%
%%%%%%%%%%%%%%%%%%%%%%%%%%%%%%%%%%%%%%%%%%%%%%%

% Authors are individuals who have significantly contributed to the
% research and preparation of the article. Group authors are allowed, if
% each author in the group is separately identified in an appendix.)

% List authors by first name or initial followed by last name and
% separated by commas. Use \affil{} to number affiliations, and
% \thanks{} for author notes.
% Additional author notes should be indicated with \thanks{} (for
% example, for current addresses).

% Example: \authors{A. B. Author\affil{1}\thanks{Current address, Antartica}, B. C. Author\affil{2,3}, and D. E.
% Author\affil{3,4}\thanks{Also funded by Monsanto.}}

\authors{Xin An\affil{1}, Shaosui Xu\affil{2}, Vassilis Angelopoulos\affil{1}, Terry Z. Liu\affil{3}, Andrew~R.~Poppe\affil{2}, Jasper~S.~Halekas\affil{4}, Ferdinand Plaschke\affil{5}}

% \affiliation{1}{First Affiliation}
% \affiliation{2}{Second Affiliation}
% \affiliation{3}{Third Affiliation}
% \affiliation{4}{Fourth Affiliation}

% \affiliation{=number=}{=Affiliation Address=}
%(repeat as many times as is necessary)
\affiliation{1}{Department of Earth, Planetary, and Space Sciences, University of California, Los Angeles, CA, 90095, USA}
\affiliation{2}{Space Sciences Laboratory, University of California, Berkeley, Berkeley, CA, 94720, USA}
\affiliation{3}{Shandong Key Laboratory of Space Environment and Exploration Technology, Institute of Space Sciences, School of Space Science and Technology, Shandong University, Shandong, China.}
\affiliation{4}{Department of Physics and Astronomy, University of Iowa, Iowa City, IA, 52242, USA}
\affiliation{5}{Institute of Geophysics and Extraterrestrial Physics, Technische Universit\"at Braunschweig, Braunschweig, Germany}

% Corresponding author mailing address and e-mail address:

% (include name and email addresses of the corresponding author.  More
% than one corresponding author is allowed in this LaTeX file and for
% publication; but only one corresponding author is allowed in our
% editorial system.)

% Example: \correspondingauthor{First and Last Name}{email@address.edu}

\correspondingauthor{Xin An}{phyax@ucla.edu}

\begin{keypoints}
\item The Hamiltonian inversion method infers lunar wake electric potentials from electron phase space densities via Vlasov equilibrium.
\item A domain decomposition strategy handles strahl-driven asymmetry on both sides of the wake and flat-top electrons near central shocks.
\item PIC simulation validation and ARTEMIS application demonstrate accurate recovery of lunar wake potentials at two evolutionary stages.
\end{keypoints}

%%%%%%%%%%%%%%%%%%%%%%%%%%%%%%%%%%%%%%%%%%%%%%%
%
%  ABSTRACT and PLAIN LANGUAGE SUMMARY
%
% A good Abstract will begin with a short description of the problem
% being addressed, briefly describe the new data or analyses, then
% briefly states the main conclusion(s) and how they are supported and
% uncertainties.

% The Plain Language Summary should be written for a broad audience,
% including journalists and the science-interested public, that will not have 
% a background in your field.
%
% A Plain Language Summary is required in GRL, JGR: Planets, JGR: Biogeosciences,
% JGR: Oceans, G-Cubed, Reviews of Geophysics, and JAMES.
% see http://sharingscience.agu.org/creating-plain-language-summary/)
%
%%%%%%%%%%%%%%%%%%%%%%%%%%%%%%%%%%%%%%%%%%%%%%%

%% \begin{abstract} starts the second page

\begin{abstract}
% [ enter your Abstract here ]
Inferring electric potentials from electron phase space density measurements in the lunar wake is complicated by two challenges: the asymmetry between the sunward and anti-sunward sides of the wake driven by the solar wind strahl, and the presence of ion acoustic shocks in the central wake. We develop the Hamiltonian inversion method, which infers the full spatial electric potential profile by exploiting the quasi-static Vlasov equilibrium condition $f = f(H)$, where $H$ is the electron Hamiltonian. The method addresses both challenges through a domain-decomposition strategy: on the two sides of the wake the potential is inferred independently by minimizing the misfit between the observed phase space density and a self-consistently reconstructed $f_\mathrm{interp}(\tilde{H})$, while in the central wake where flat-top trapped electron distributions are present the potential is inferred directly from the flat-top width. We validate the method against particle-in-cell simulation data at two evolutionary stages of the lunar wake: an early stage where strahl asymmetry is strong but no shocks have formed, and a later stage where ion acoustic shocks and flat-top distributions are present. We then apply the method to two ARTEMIS lunar wake crossings at the same evolutionary stages, inferring normalized potential drops of $e\Delta\varphi/T_e \sim 15$ and $\sim 5$ respectively and capturing shock-associated potential enhancements in the central wake. The method is broadly applicable to plasma environments where electrons are in quasi-static equilibrium with a field-aligned electric potential.
\end{abstract}

\section*{Plain Language Summary}
% Enter your Plain Language Summary here or delete this section.
% Here are instructions on writing a Plain Language Summary: 
% https://www.agu.org/Share-and-Advocate/Share/Community/Plain-language-summary
The Moon has no atmosphere or global magnetic field, allowing the solar wind, a continuous stream of charged particles from the Sun, to hit its surface directly and create an empty region behind the Moon called the lunar wake. Electric fields inside this wake control how it refills with plasma, but they are too weak to measure directly with spacecraft instruments. Instead, we infer them from electron energy measurements. Existing methods struggle in the lunar wake because the Sun emits a beam of energetic electrons in one direction, making the two sides of the wake asymmetric, and because colliding plasma streams near the wake center generate shocks that further complicate the electron distributions. We developed a new method that handles both challenges by dividing the wake into regions and analyzing each one separately according to its local conditions. We validated the method using computer simulations and applied it to measurements from the ARTEMIS spacecraft orbiting the Moon, successfully recovering the electric potential structure including the asymmetry and shock-associated features.

%%%%%%%%%%%%%%%%%%%%%%%%%%%%%%%%%%%%%%%%%%%%%%%
%
%  BODY TEXT
%
%%%%%%%%%%%%%%%%%%%%%%%%%%%%%%%%%%%%%%%%%%%%%%%

%%% Suggested section heads:
% \section{Introduction}
%
% The main text should start with an introduction. Except for short
% manuscripts (such as comments and replies), the text should be divided
% into sections, each with its own heading.

% Headings should be sentence fragments and do not begin with a
% lowercase letter or number. Examples of good headings are:

% \section{Materials and Methods}
% Here is text on Materials and Methods.
%
% \subsection{A descriptive heading about methods}
% More about Methods.
%
% \section{Data} (Or section title might be a descriptive heading about data)
%
% \section{Results} (Or section title might be a descriptive heading about the
% results)
%
% \section{Conclusions}

\section{Introduction}
%Text here ===>>>
% A section at the beginning why to I need to study the relevant aspect of space physics. You must motivate the publication of this technical advancement in JGR Space Physics by convincing readers that the science area to which it pertains is interesting.
% Structure of lunar wake and lunar wake potential
The Moon lacks a global intrinsic magnetic field and a substantial atmosphere, allowing the solar wind to impinge directly on its surface. This absorption creates a plasma void immediately downstream of the Moon, which the solar wind subsequently refills both perpendicular and parallel to the interplanetary magnetic field, forming an extended region known as the lunar wake. Here we focus on the parallel refilling process. Early in this process, parallel electric fields are governed by gradients in electron density and temperature \cite<e.g.,>[]{gurevich1966self,crow1975expansion,denavit1979collisionless,mora2003plasma,halekas2014effects}. As refilling progresses, counter-streaming supersonic ion beams from opposite flanks of the wake converge near its center, driving ion acoustic shocks that come to dominate the parallel electric field \cite{farrell1998simple,birch2001detailed,halekas2014first,malaspina2019properties,An2025plasma,liu2025artemis}. The resulting electric potential thus both governs field-aligned particle motion and is itself shaped by the collective dynamics of those particles.

Understanding the electric potential structure of the lunar wake has implications beyond the Moon itself. Similar plasma void and refilling dynamics arise downstream of other unmagnetized or weakly magnetized bodies \cite{halekas2015moon,kivelson2016moons}, including asteroids, comets, and the moons of outer planets. More broadly, the lunar wake serves as a natural laboratory for studying fundamental plasma processes, including electrostatic shock formation, electron trapping by nonlinear Landau resonance, and the interplay between field-aligned particle transport and self-consistent electric fields, in a relatively accessible and well-instrumented environment.

% Difficulty with direct measurements; inference of lunar wake potentials from electron phase space densities (Halekas & Xu)
Considering a potential drop of order $\sim 100$\,V from the solar wind to the wake center \cite{xu2019mapping}, the corresponding average parallel electric field is on the order of $\sim 100\,\mathrm{V}/R_l \approx 0.06$\,mV/m (where $R_l = 1737$\,km is the lunar radius), well below both the motional electric field of a few mV/m in the solar wind at lunar orbit and the sensitivity threshold of existing electric field instruments on spacecraft \cite<e.g.,>[]{Bonnell08}. Direct measurement of wake potentials is therefore generally not feasible, with the exception of localized, intense potential enhancements associated with electrostatic shocks that develop in the central wake beyond a certain radial distance from the Moon. As an alternative, lunar wake potentials have been inferred from electron phase space density measurements under the assumption that electron distributions are in quasi-static equilibrium with the local electric potential. Leveraging this approach, \citeA{halekas2014first} and \citeA{xu2019mapping} compared field-aligned electron distribution functions inside the wake against reference measurements taken outside the wake, identifying systematic energy shifts between the parallel and antiparallel electron spectra across the wake boundary. These spectral energy shifts are then interpreted as potential drops between the reference location and points within the wake interior. Hereafter we refer to this technique as the Hamiltonian shift method. While practical and widely applied, the Hamiltonian shift method faces fundamental limitations in the lunar wake, motivating the development of the Hamiltonian inversion method presented in this work.

% Drawback of previous methods; What is our new contribution.
The Hamiltonian shift method requires a minimum parallel energy to ensure that the electron population under comparison can penetrate through the entire wake \cite{halekas2005electrons,halekas2014first,xu2019mapping}. This requirement implicitly assumes that open trajectories exist in phase space $(x, p_x)$ connecting any two points along $x$ (where $x$ is the coordinate along the magnetic field and across the wake and $p_x$ the corresponding electron momentum), and that the phase space densities along these trajectories are above the measurement noise floor. These two conditions are not always simultaneously satisfied. For instance, immediately downstream of the Moon, the parallel energy required to traverse the potential well may be so large that the corresponding phase space densities fall below instrument sensitivity. Furthermore, the Hamiltonian shift method does not utilize reflected electrons, whose phase space densities are often well measured and carry valuable information about the shape of the electric potential. To address these limitations, we develop the Hamiltonian inversion method, which incorporates both penetrating and reflected electrons and is applicable to arbitrary potential profiles including those containing electrostatic shocks. The method is grounded in the quasi-static Vlasov equilibrium condition $f = f(H)$ and employs a domain-decomposition strategy to handle the physical complexity of the central wake, where counter-streaming ion beams generate shocks and trap electrons. We describe the Hamiltonian inversion method in Section~\ref{sec:method}, validate it against simulation data in Section~\ref{sec:validate}, apply it to spacecraft observations in Section~\ref{sec:examples}, and summarize our findings in Section~\ref{sec:conclusion}.

\section{Methodology}\label{sec:method}
The electron Hamiltonian along magnetic field lines is
\begin{linenomath}
    \begin{align}
    H(x, p_x, \mu, t) = \frac{p_x^2}{2 m_e} + \mu B(x) - e \varphi(x,t),
    \end{align}
\end{linenomath}
where $e$ is the elementary charge, $m_e$ is the electron mass, $\varphi(x,t)$ is the electric potential, $B(x)$ is the magnetic field magnitude along the field line, and $\mu = p_\perp^2 / (2 m_e B)$ is the electron magnetic moment. The magnetic moment $\mu$ is an adiabatic invariant provided that $B(x)$ varies slowly compared to the gyroperiod. When $B$ varies along the field line, $\mu$ serves as an additional phase-space coordinate, and the distribution function becomes $f(x, p_x, \mu)$, rendering the inversion problem three-dimensional. In this work, we consider cases in which variations in $B$ are negligible: the simulations employ a uniform background magnetic field, and we select spacecraft observations where $B$ remains sufficiently steady along the spacecraft trajectory. Under this assumption, the $\mu B(x)$ term becomes constant and can be omitted, so that the Hamiltonian reduces to
\begin{linenomath}
\begin{align}
H(x, p_x, t) = \frac{p_x^2}{2 m_e} - e \varphi(x,t).
\end{align}
\end{linenomath}
Accordingly, the distribution function depends only on $(x, p_x)$, and the inversion is carried out in the corresponding two-dimensional phase space, as described below. We further assume that electron parallel dynamics are primarily governed by $\varphi(x,t)$, neglecting other processes such as pitch-angle scattering. The evolution of electron phase space density follows the Vlasov equation $\partial_t f + \{f, H\} = 0$, where $\{\cdot,H\}$ denotes the Poisson bracket. In the lunar wake, both self-similar expansion and electrostatic shocks evolve on the slow ion timescale ($\omega_{pi}^{-1}$), while electrons respond on the fast electron timescale ($\omega_{pe}^{-1}$). This separation ensures electrons maintain quasi-equilibrium at any instant ($\partial_t f \sim 0$), such that the electron phase space density depends only on the Hamiltonian:
\begin{linenomath}
    \begin{align}\label{eq:f=fH}
        f = f(H).
    \end{align}
\end{linenomath}
This relationship provides the constraint for inferring $\varphi$ from the observed $f$. However, applying this constraint across the full domain faces two challenges specific to the lunar wake. First, the solar wind strahl, a field-aligned beam of suprathermal electrons propagating away from the Sun, renders the electron distribution inherently asymmetric between the sunward and anti-sunward sides of the wake, making $f(H)$ multi-valued when evaluated globally. Second, when plasmas from the two sides encounter each other near the wake's center, ion acoustic shocks form and trap electrons through nonlinear Landau resonance, producing flat-top distributions for which $f(H)$ is ill-defined. Our Hamiltonian inversion method addresses both challenges through a domain-decomposition strategy, described in the following steps.

\begin{figure}[tphb]
    \centering
    \includegraphics[width=\linewidth]{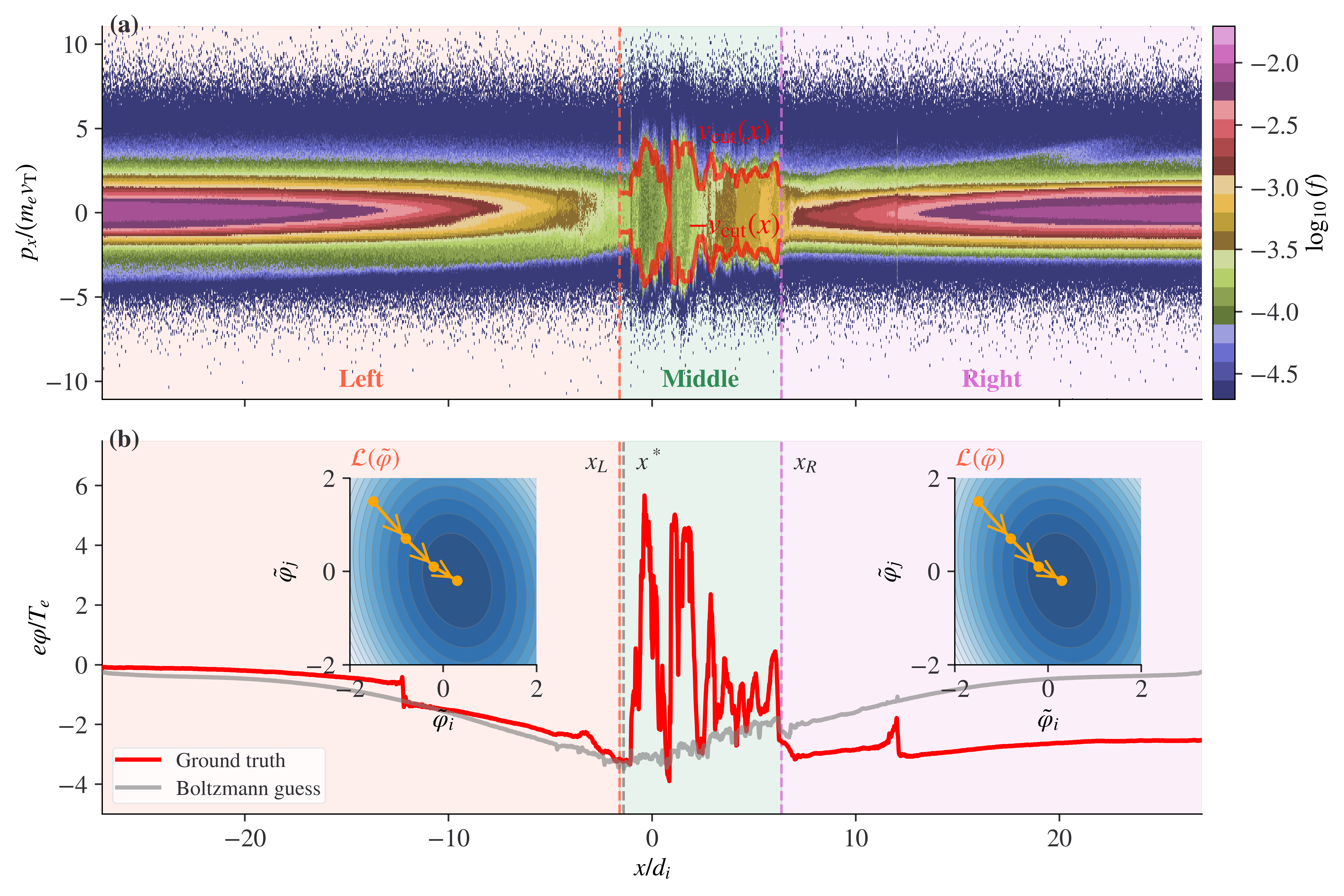}
    \caption{Schematic illustration of the Hamiltonian inversion method. (a) Electron phase space density $f(x, p_x)$ with three spatial domains indicated by shading: left (red), middle (green), and right (purple). The red contour marks the separatrix that bounds the trapped electron population in the middle region. (b) Electric potential $e\varphi/T_e$ from the ground-truth (red) and the Boltzmann initial guess (gray). The left and right insets illustrate the working principle of the L-BFGS-B optimization: starting from the Boltzmann initial guess $\tilde{\varphi}_\mathrm{init}$, the algorithm iteratively minimizes the error functional $\mathcal{L}(\tilde{\varphi})$ (schematic contours) along a quasi-Newton descent trajectory (orange arrows) until convergence to the optimal solution $\tilde{\varphi}_\mathrm{opt}$. This optimization is applied independently to the left and right domains. In the middle region, the potential is inferred directly from the flat-top width via Equation \eqref{eq:phi_middle}. Vertical dashed lines mark the domain boundaries $x_L$, $x^*$, and $x_R$.}
    \label{fig:sketch}
\end{figure}

\textit{Step 1: Initial guess and domain split point.}
We aim to find the potential $\varphi(x)$ that best satisfies $f = f(H)$ for the measured electron distributions. Our initial guess is motivated by the Boltzmann distribution:
\begin{linenomath}
    \begin{align}\label{eq:phi-init}
        \tilde{\varphi}(x) = \frac{T_e}{e}\ln \left(\frac{n(x)}{n_{0}} \right),
    \end{align}
\end{linenomath}
where $n_0$ and $T_e$ are the reference electron density and temperature in the solar wind, respectively, and $n(x) = \int \mathrm{d}p_x\, f(x,p_x)$ is the measured electron density at position $x$. The Boltzmann estimate captures the large-scale structure of the potential well reliably enough to identify the preliminary domain split point $x^* = \mathrm{argmin}_x\, \tilde{\varphi}(x)$ [Figure~\ref{fig:sketch}(b)], which is used in the domain decomposition of Step 2.

\textit{Step 2: Domain decomposition.}
The potential minimum $x^*$ serves as the preliminary split point between the left ($x < x^*$) and right ($x > x^*$) domains, motivated by the fact that the separatrix between trapped and passing electron orbits is anchored to the potential minimum. The middle region, where flat-top distributions are present, is identified by examining the effective electron thermal velocity,
\begin{linenomath}
    \begin{align}
        v_\mathrm{th}(x) = \sqrt{\frac{\int \mathrm{d}p_x\, f(x,p_x)\, p_x^2 / m_e^2}{\int \mathrm{d}p_x\, f(x,p_x)}},
    \end{align}
\end{linenomath}
as a function of position. Electron heating by the ion acoustic shocks produces a sharp, localized enhancement of $v_\mathrm{th}(x)$ near $x^*$, distinct from the gradual large-scale variation of $v_\mathrm{th}(x)$ associated with inter-penetrating populations. The middle region is defined as the spatially connected region near $x^*$ where $v_\mathrm{th}(x)$ normalized by its local background on each side exceeds a threshold factor $\alpha_\mathrm{th}$, with the boundary extended further outward as long as $|dv_\mathrm{th}/dx|$ remains large, capturing the full shock transition. The resulting domain boundaries $x_L$ and $x_R$ partition the domain into three regions: left ($x < x_L$), middle ($x_L \leq x \leq x_R$), and right ($x > x_R$) [Figure~\ref{fig:sketch}(a)]. When no middle region is detected, as is the case in the early stage of wake formation before ion acoustic shocks develop, the algorithm reduces to a two-domain inversion with $x_L = x_R = x^*$.

\textit{Step 3: Left and right domain inversions.}
The left and right domains are inverted independently using the following procedure. For a given $\tilde{\varphi}(x)$ on the domain, each phase space pixel $(x, p_x)$ is assigned a Hamiltonian value
\begin{linenomath}
    \begin{align}
        \tilde{H}(x, p_x) = \frac{p_x^2}{2 m_e} - e \tilde{\varphi}(x).
    \end{align}
\end{linenomath}
Under the assumption $f = f(H)$, the phase space density should depend only on $\tilde{H}$. To reconstruct this functional dependence, we bin the set of pairs $\{(\tilde{H}(x, p_x),\, f(x, p_x))\}$ across all phase space pixels within the domain. The number of bins is chosen as a balance between two competing requirements: too few bins fail to resolve the structure of $f(\tilde{H})$, while too many bins leave insufficient counts per bin for reliable statistics. In practice, we use $N_\mathrm{bin} = 150$ bins of equal width in $\tilde{H}$. Only bins containing at least $M_{\min} = 50$ pixels are retained; bins below this threshold are excluded from the interpolation to avoid contamination by poorly sampled regions of phase space. The median phase space density $\tilde{f}_k$ and mean Hamiltonian $\tilde{H}_k$ are computed within each valid bin $k$. The predicted phase space density $f_\mathrm{interp}(\tilde{H})$ is then obtained by linearly interpolating over the pairs $\{(\tilde{H}_k,\, \tilde{f}_k)\}$. We then define the error functional
\begin{linenomath}
    \begin{align}
        \mathcal{L}(\tilde{\varphi}) = \sum_{\substack{x,\, p_x \\ f > f_\mathrm{floor} \\ \tilde{H} \in [\tilde{H}_\mathrm{min}, \tilde{H}_\mathrm{max}]}} \left[ \ln f(x, p_x) - \ln f_\mathrm{interp}(\tilde{H}(x, p_x)) \right]^2,
    \end{align}
\end{linenomath}
where the sum is restricted to pixels with $f$ above a noise floor $f_\mathrm{floor}$ and with $\tilde{H}$ within the statistically reliable range $[\tilde{H}_\mathrm{min}, \tilde{H}_\mathrm{max}]$ spanned by the valid bins. Pixels outside this range are excluded because $f_\mathrm{interp}$ is poorly constrained there due to insufficient statistics. The use of logarithms ensures that the minimization treats errors multiplicatively, giving equal weight to pixels across the many decades of dynamic range typical in electron phase space densities. We minimize $\mathcal{L}(\tilde{\varphi})$ using the L-BFGS-B algorithm [see \citeA{zhu1997algorithm} and Figure~\ref{fig:sketch}(b)], a limited-memory quasi-Newton method that approximates the inverse Hessian from a history of gradient evaluations, with gradients computed numerically via finite differences. At each iteration of the optimizer, $\tilde{H}(x, p_x)$ is recomputed from the updated $\tilde{\varphi}$, and the binning and interpolation are repeated to obtain a self-consistent $f_\mathrm{interp}(\tilde{H})$. To avoid local minima and improve convergence, we adopt a three-level multigrid approach. First, $\tilde{\varphi}$ is solved on a coarse spatial grid of $\sim$100 points, with each level run for a fixed maximum number of iterations. The solution is then interpolated onto a medium-resolution grid of $\sim$500 points and used as the initial guess for a second optimization. Finally, the solution is interpolated onto the full-resolution grid matching the spatial resolution of the phase space density measurements, where the optimization is performed for the last time. This coarse-to-fine strategy ensures that the large-scale structure of $\tilde{\varphi}$ is captured first, guiding the optimizer toward the correct solution at finer scales. This procedure resolves the multi-valued $f(H)$ problem caused by the strahl asymmetry: on each side separately, $f(H)$ is well-defined and approximately single-valued, since the strahl contribution is unidirectional and the local electron population is in quasi-equilibrium within its half-domain. Any penetrating electrons from the opposite side carry very low phase space densities and contribute negligibly to the binned $f_\mathrm{interp}$.

\textit{Step 4: Middle region inversion via flat-top width.}
In the middle region, the presence of ion acoustic shocks and nonlinear Landau trapping renders $f(H)$ ill-defined  [Figure~\ref{fig:sketch}], precluding the use of the $f_\mathrm{interp}$-based optimization. Instead, we exploit the flat-top structure of the trapped electron velocity distribution \cite{An2025plasma,liu2025artemis}. Electrons trapped in the potential well have bounced many times and uniformly populate the accessible phase space, producing a distribution that is approximately constant in $f$ up to the cutoff velocity and drops sharply beyond it. The cutoff velocity $v_\mathrm{cut}(x)$ is directly observable as the velocity at which $f(x, p_x)$ drops below a fraction $\alpha = 0.5$ of its plateau value $f_0(x) = \max_{p_x} f(x, p_x)$ [Figure~\ref{fig:sketch}(a)]. The potential in the middle region is then directly inferred as \cite{An2025plasma}
\begin{linenomath}
    \begin{align}\label{eq:phi_middle}
        \tilde{\varphi}(x) = \tilde{\varphi}(x_L) + \frac{m_e}{2e}\left[v_\mathrm{cut}^2(x) - v_\mathrm{cut}^2(x_L)\right],
    \end{align}
\end{linenomath}
anchored to the converged left-domain potential at the left boundary $x_L$. This direct inversion is robust precisely where the $f(H)$ fitting fails: the flat-top edge is a sharp, well-defined feature in velocity space that can be reliably detected even in the presence of multiple shocks and interleaved populations. Crucially, since the flat-top electrons have sampled the potential well on both sides through repeated bouncing, $v_\mathrm{cut}(x)$ encodes the local potential relative to the potential minimum, providing the only reliable constraint on the potential drop across the middle region.

\textit{Step 5: Stitching.}
The three solutions are combined into a single $\tilde{\varphi}(x)$ profile. The left solution is used directly for $x < x_L$. The middle solution, anchored to $\tilde{\varphi}(x_L)$ by construction via Equation~\eqref{eq:phi_middle}, is used for $x_L \leq x \leq x_R$. The right solution is shifted by a constant offset $\Delta\tilde{\varphi} = \tilde{\varphi}_\mathrm{middle}(x_R) - \tilde{\varphi}_\mathrm{right}(x_R)$ to enforce continuity at the right boundary, yielding a globally continuous potential profile.

For comparison, we also apply the Hamiltonian shift method \cite{halekas2005electrons,halekas2014first,xu2019mapping} to the same data. In this method, the electric potential at each position $x$ is inferred by identifying the energy shift $\Delta H(x)$ that best aligns the local electron distribution $f(x, p_x)$ with a reference distribution $f_\mathrm{ref}(p_x)$ measured in the ambient solar wind, under the assumption that $f(p_x^2/2m_e + \Delta H) = f_\mathrm{ref}(p_x^2/2m_e)$ for electrons on open trajectories, giving $e\varphi(x) = \Delta H(x)$. Two independent solutions are obtained using electrons traveling in opposite directions: $\varphi_l$, derived from electrons with $p_x > 0$ referenced to the left boundary, and $\varphi_r$, derived from electrons with $p_x < 0$ referenced to the right boundary. A composite potential $\varphi_\mathrm{composite}$ is constructed by matching the two solutions at a midpoint $x_\mathrm{mid}$: $\varphi_\mathrm{composite}(x < x_\mathrm{mid}) = \varphi_l(x < x_\mathrm{mid})$ and $\varphi_\mathrm{composite}(x > x_\mathrm{mid}) = \varphi_r(x > x_\mathrm{mid}) + \varphi_\mathrm{shift}$, where $\varphi_\mathrm{shift} = \varphi_l(x_\mathrm{mid}) - \varphi_r(x_\mathrm{mid})$ enforces continuity. Ideally $x_\mathrm{mid}$ should be the location of the global potential minimum, which is not known a priori; here we manually select $x_{\mathrm{mid}}$, which is approximately the location of minimum plasma density.
% For both PIC simulations, $x_{\mathrm{mid}}/d_i = 5$ is used for simplicity; the potential difference $\varphi_{\mathrm{shift}}$ is $-9.4 eU/T_e$ and $-1.8 eU/T_e$, respectively, for the two cases. 

It is worth pointing out an important difference between the Hamiltonian inversion and Hamiltonian shift methods in the middle region. In this region, electron velocity distributions exhibit a flat-top signature: electrons from the two sides undergo nonlinear trapping and mix to approximately constant phase space densities within the trapping island. The Hamiltonian inversion method exploits the fact that the width of the trapping island directly encodes the local electric potential, inferring the latter from the former via Equation~\eqref{eq:phi_middle}. The Hamiltonian shift method, on the other hand, relies on the energy shift of passing electrons outside the trapping island, which requires high-energy electrons whose phase space densities are often low and whose two populations ($p_x > 0$ and $p_x < 0$) must be stitched together at a carefully chosen location to obtain a physically consistent solution. The domain-decomposition strategy of the Hamiltonian inversion method naturally resolves this stitching problem by identifying the middle region explicitly and treating it with a physically motivated direct inversion, rather than relying on the continuity of passing electron populations across the trapping region.

\section{Validation}\label{sec:validate}
We validate the Hamiltonian inversion method against data from Vector Particle-in-Cell (VPIC) simulation \cite{bird2021vpic,bowers2008ultrahigh,bowers20080,bowers2009advances} of the lunar wake, where electron phase space density $f(x, p_x)$ can be directly constructed from particles, and the true electric potential is known exactly from the simulation fields and serves as the ground truth for comparison.

We model a one-dimensional slice extending across the lunar wake as it is convected by the solar wind, so that the simulation's temporal evolution maps to the wake's radial evolution through $r = v_\mathrm{sw} \cdot t$, where $r$ is the radial distance from the Moon, $v_\mathrm{sw}$ is the solar wind velocity, and $t$ is the simulation time \cite{birch2001detailed,birch2001particle}. The domain spans $-30\,d_i \leq x \leq 30\,d_i$, with the interplanetary magnetic field oriented in the $+x$ direction. The initial density profile places solar wind plasma at $R_l \leq |x| \leq 30\,d_i$ and a vacuum at $|x| < R_l$, where the lunar radius is $R_l = 13.2\,d_i$. We use a reduced ion-to-electron mass ratio $m_i/m_e = 100$ to make the computational cost affordable. The electron velocity distribution is initialized as a three-component model comprising a Maxwellian core ($0.9n_0$ with $n_0$ being the total electron density), a kappa halo ($0.06 n_0$), and a field-aligned kappa strahl ($0.04 n_0$) \cite{vstverak2009radial,maksimovic2000solar}. The strahl component introduces the sunward/anti-sunward asymmetry in the electron distribution that is the primary motivation for the domain decomposition in the Hamiltonian inversion method. The simulation time maps approximately to radial distance downstream of the Moon as $r/R_l = 5.5 \times 10^{-4}\,t\,\omega_{pi}$ \cite{An2025plasma}. Further details of the simulation setup, including boundary conditions and numerical parameters, are identical to those described in our companion paper. 

To extract the ground-truth electric potential for comparison with the Hamiltonian inversion, we average the simulated electric field over a window of 200 time steps ending at the target time to reduce particle noise, and then integrate the time-averaged field along $x$ to obtain $\varphi(x)$. The synthetic electron phase space density $f(x, p_x)$ is constructed by depositing PIC particle data onto a uniform grid spanning $x \in [-30\,d_i, 30\,d_i]$ with 3000 spatial cells and $p_x \in [-18.5\,m_e v_{\mathrm{T}}, 18.5\,m_e v_{\mathrm{T}}]$ with 400 momentum cells, where $v_{\mathrm{T}}$ is the thermal velocity of core electrons.

\subsection{Case 1: Early Stage Without Shocks}

\begin{figure}[tphb]
    \centering
    \includegraphics[width=\linewidth]{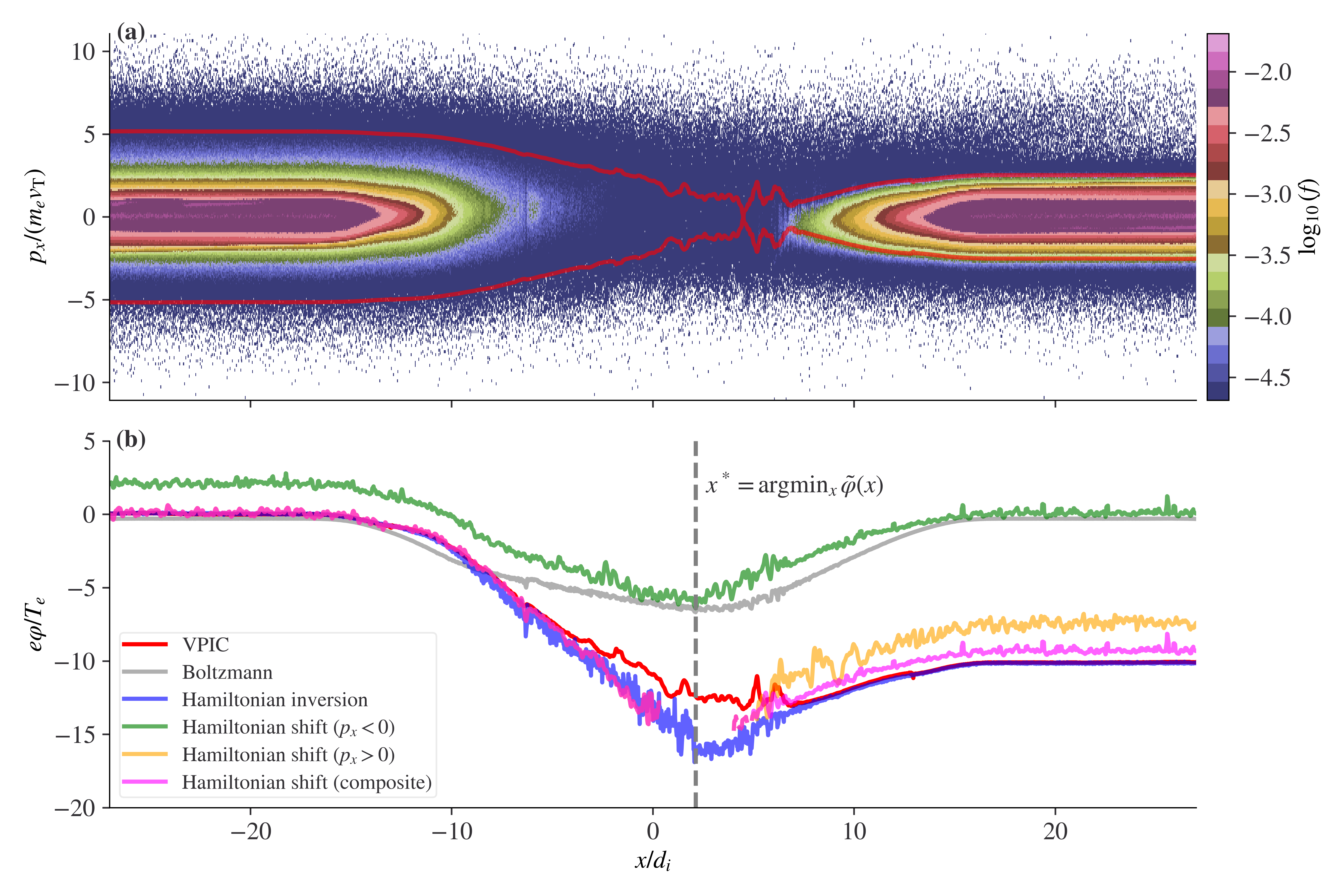}
    \caption{(a) Electron phase space density $f(x, p_x)$ at simulation time $t\,\omega_{pi} = 3000$, corresponding to a radial distance $r/R_l = 1.65$ downstream of the Moon. The asymmetry between the left and right sides of the wake is clearly visible, driven by the strahl electrons streaming freely on the right side while the left strahl has not yet penetrated through the wake. The red curve marks the separatrix, defined as the locus of points with $p_x = 0$ at the location of the global potential minimum $\varphi_\mathrm{min}$. (b) Electric potential $\varphi(x)$ inferred from the Boltzmann relation (gray), Hamiltonian inversion (blue), and Hamiltonian shift method using anti-parallel ($p_x < 0$, green) and parallel ($p_x > 0$, yellow) electrons, as well as the stitched composite potential (magenta), all benchmarked against the ground-truth PIC potential (red). The stitching point for the Hamiltonian shift method is $x_\mathrm{mid} = 5\,d_i$.}
    \label{fig:modeldata-t3000}
\end{figure}

\begin{figure}[tphb]
    \centering
    \includegraphics[width=\linewidth]{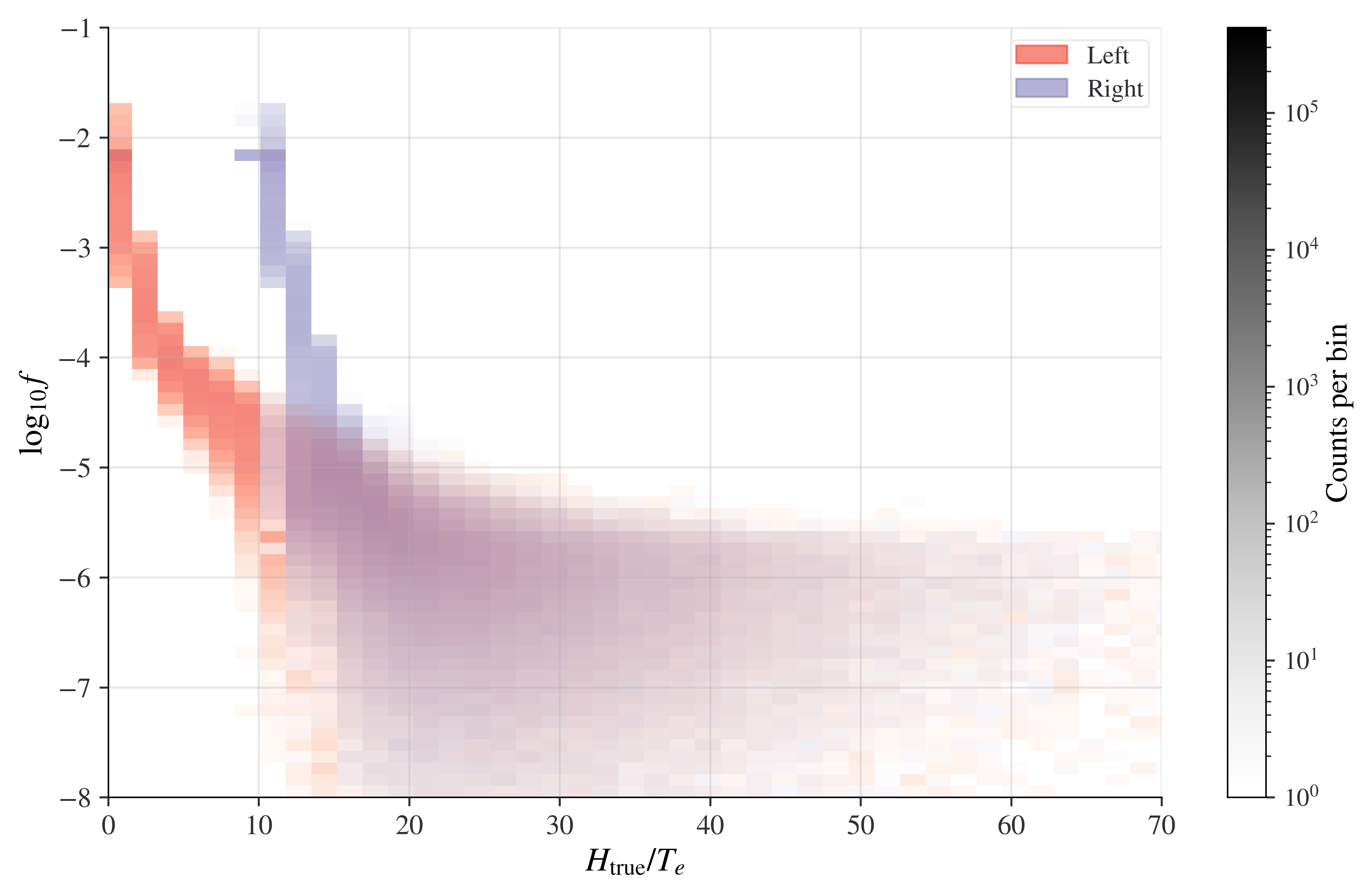}
    \caption{Distribution of electron phase space density $f$ as a function of the true Hamiltonian $H = p_x^2/(2m_e) - e\varphi_\mathrm{true}(x)$ at simulation time $t\,\omega_{pi} = 3000$, corresponding to $r/R_l = 1.65$ downstream of the Moon. The color map indicates the number of phase space pixels per $(H, \log_{10} f)$ bin, with left (red) and right (purple) domains shown separately. If $f = f(H)$ held globally, all pixels would collapse onto a single curve; instead, the two domains trace clearly distinct $f(H)$ relations, reflecting the strahl-driven asymmetry between the two sides of the wake.}
    \label{fig:f-vs-Htrue-t3000}
\end{figure}

The first case is taken from the early stage ($t\,\omega_{pi} = 3000$) of our simulation, corresponding to a location close to the Moon ($r/R_l = 1.65$) where the wake is still strongly density-depleted. At this location, the counter-streaming supersonic ion beams have not yet fully converged and no ion acoustic shocks have formed in the wake center. However, the electron distributions on the two sides of the wake are strongly asymmetric [Figure~\ref{fig:modeldata-t3000}(a)]: strahl electrons on the right side stream freely out of the domain, whereas those from the left have not yet penetrated through the wake to reach the right side. This asymmetry renders $f(H)$ multi-valued across the full domain [Figure \ref{fig:f-vs-Htrue-t3000}], and this case therefore isolates and tests the domain-decomposition strategy for handling strahl asymmetry, without the added complexity of the middle-region flat-top inversion.

\begin{figure}[tphb]
    \centering
    \includegraphics[width=\linewidth]{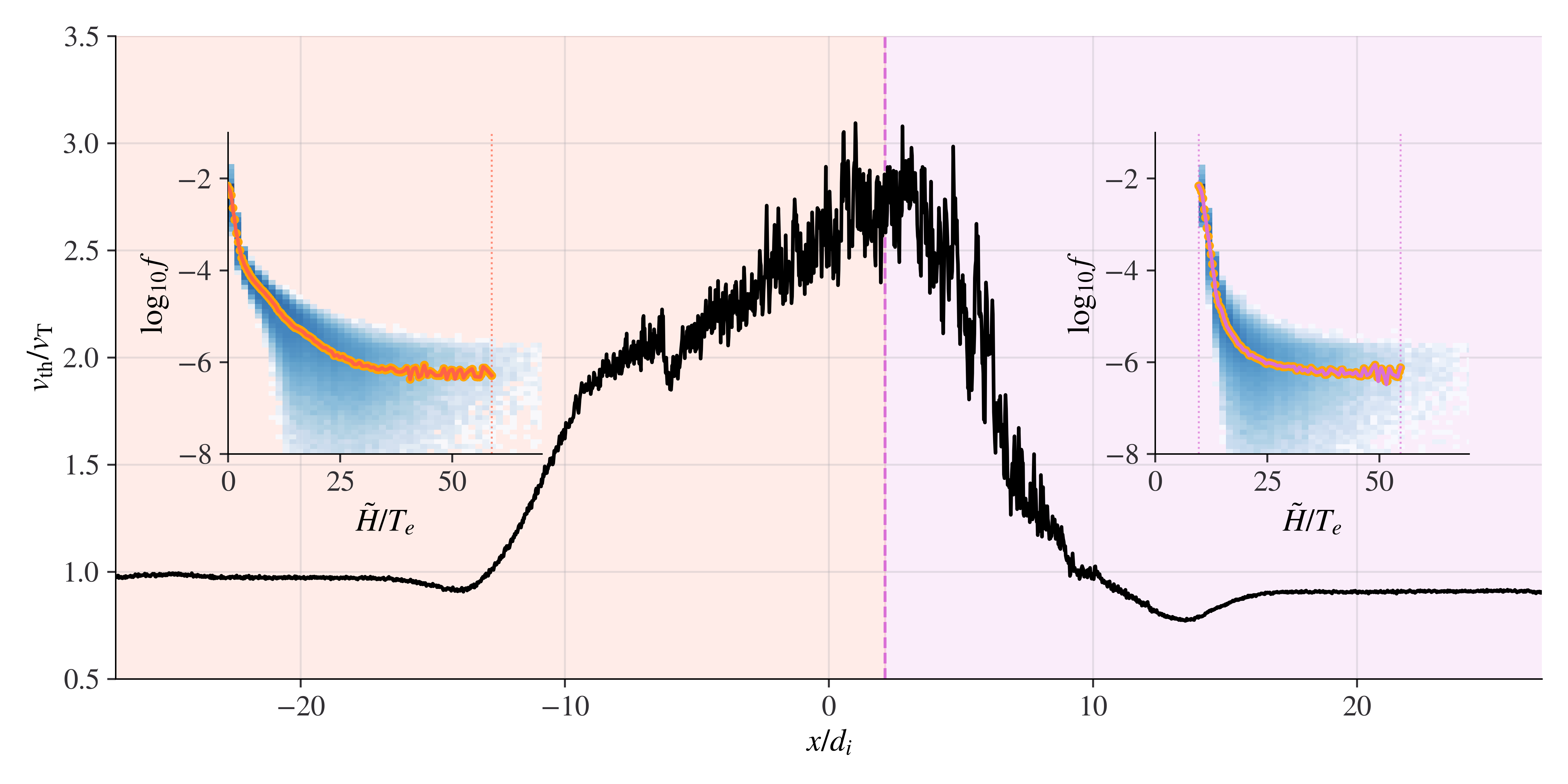}
    \caption{Domain decomposition at simulation time $t\,\omega_{pi} = 3000$, corresponding to $r/R_l = 1.65$ downstream of the Moon. The main panel shows the $v_\mathrm{th}(x)$ profile (black curve), which rises gradually in the central wake due to counter-streaming electrons but exhibits no sharp shock transition; accordingly, no middle region is detected and the domain is split into left (red shading) and right (purple shading) regions at $x^*/d_i = 2.1$, the location of the minimum of the Boltzmann potential. The left and right insets show the distribution of phase space pixels in the $(\tilde{H}, f)$ plane for each domain, with the color map indicating pixel counts per bin and the solid curve showing $f_\mathrm{interp}(\tilde{H})$. The two $f_\mathrm{interp}$ curves are clearly distinct, reflecting the strahl-driven asymmetry between the two sides of the wake and motivating the need for separate inversions.}
    \label{fig:vth-t3000}
\end{figure}

Figure~\ref{fig:vth-t3000} shows the domain decomposition. The domain is split at $x^*/d_i = 2.1$, the location of the minimum of the Boltzmann potential [see Equation~\eqref{eq:phi-init} and the gray curve in Figure~\ref{fig:modeldata-t3000}(b)], which coincides with the density minimum. Near $x^*$ in the wake center, $v_\mathrm{th}(x)$ rises gradually due to counter-streaming electrons, with no sharp shock transition detected; accordingly, no middle region is identified and the algorithm reduces to a two-domain inversion. The two insets in Figure~\ref{fig:vth-t3000} show the distinct $f(\tilde{H})$ relations reconstructed independently for the left and right domains, highlighting the need for separate inversions. In particular, the core electrons, which carry the highest phase space densities and thus the most reliable statistics, exhibit a systematic shift in $\tilde{H}$ between the two domains, directly reflecting the asymmetry in $\tilde{\varphi}(x)$ driven by the strahl.

Figure~\ref{fig:modeldata-t3000}(b) compares the inferred potential with the ground truth. Starting from the Boltzmann potential as the initial guess, the Hamiltonian inversion method [the blue curve in Figure~\ref{fig:modeldata-t3000}(b)] most accurately captures the asymmetry of $\varphi(x)$ on the two sides of the wake. A slight potential enhancement near the wake center, associated with electron heating by ion acoustic waves, is not fully recovered; this is expected since no middle-region inversion is applied at this early stage.

The Hamiltonian shift method reproduces both the asymmetric potential structure across the wake and the magnitude of the potential drop, consistent with the Hamiltonian inversion method [Figure~\ref{fig:modeldata-t3000}(b)]. However, it fails near the deepest part of the wake center, where the minimum kinetic energy required for electrons to traverse the full potential well is large and the phase space densities along these open trajectories suffer from low counting statistics.
% In contrast, the Hamiltonian shift method underestimates the potential gradient in the wake center. \textcolor{red}{[Todo (Shaosui): assessment of Hamiltonian shift method performance.]}

\subsection{Case 2: Late Stage With Shocks}

\begin{figure}[tphb]
    \centering
    \includegraphics[width=\linewidth]{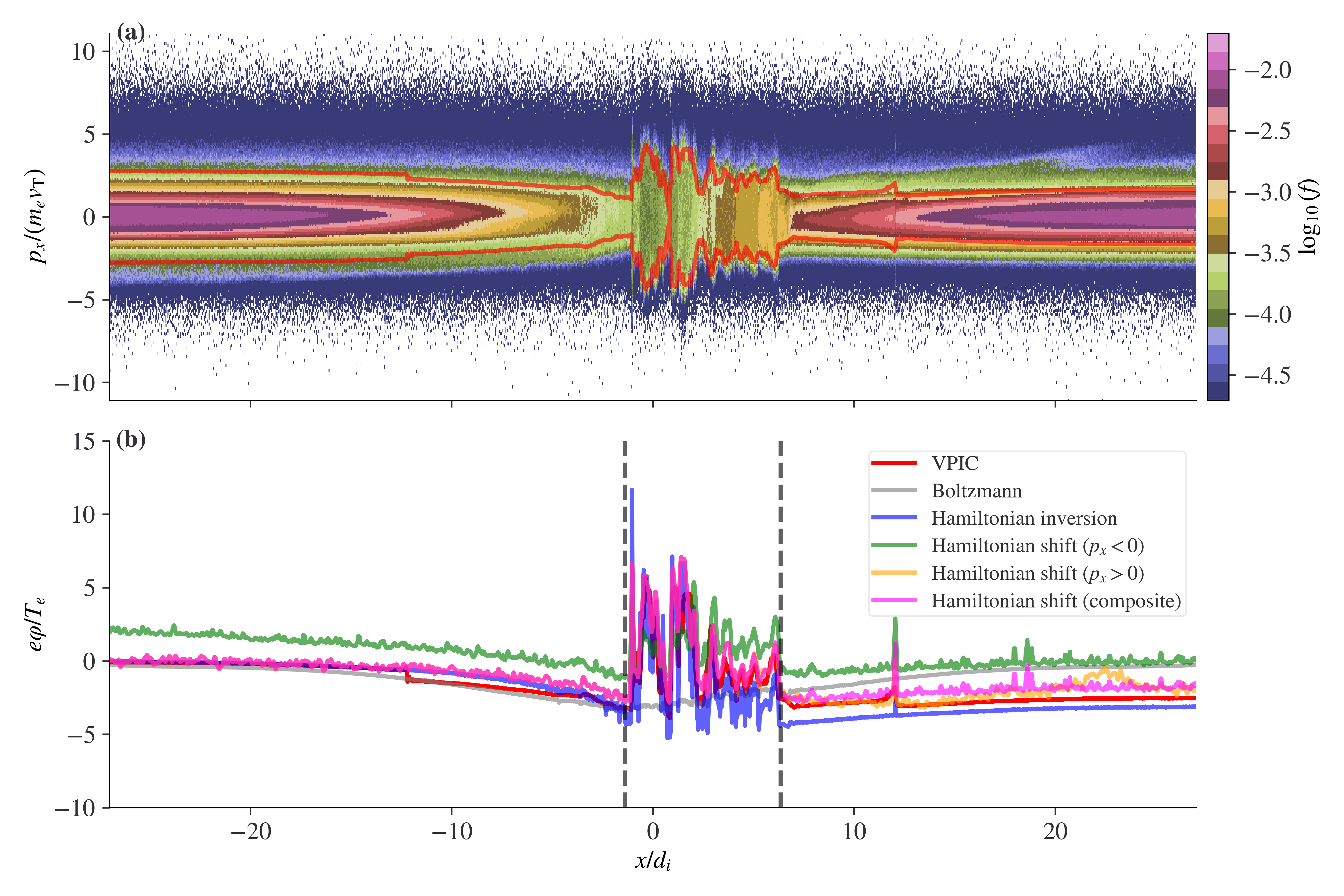}
    \caption{(a) Electron phase space density $f(x, p_x)$ at simulation time $t\,\omega_{pi} = 10000$, corresponding to a radial distance $r/R_l = 5.5$ downstream of the Moon. Ion acoustic shocks have developed in the central wake, and flat-top electron distributions are clearly visible in the middle region between the two shocks. The red curve marks the separatrix, defined as the locus of points with $p_x = 0$ at the location of the global potential minimum $\varphi_\mathrm{min}$. (b) Electric potential $\varphi(x)$ inferred from the Boltzmann relation (gray), Hamiltonian inversion (blue), and Hamiltonian shift method using anti-parallel ($p_x < 0$, green) and parallel ($p_x > 0$, yellow) electrons, as well as the stitched composite potential (magenta), all benchmarked against the ground-truth PIC potential (red). Vertical dashed lines mark the detected middle region boundaries $x_L$ and $x_R$. The stitching point for the Hamiltonian shift method is $x_\mathrm{mid} = 5\,d_i$. The two sharp discontinuities in the ground-truth potential near $x = \pm 13\,d_i$ correspond to the lunar surface boundary at $R_l = 13.2\,d_i$.}
    \label{fig:modeldata-t10000}
\end{figure}

The second case is taken from the late stage ($t\,\omega_{pi} = 10000$) of our simulation, corresponding to a location further downstream ($r/R_l = 5.5$) where counter-streaming ion beams have converged and ion acoustic shocks have developed near the wake center. Electrons are trapped by the shock potential through nonlinear Landau resonance and exhibit flat-top distributions in the central wake [see Figure~\ref{fig:modeldata-t10000}(a) and \citeA{An2025plasma}]. This case tests the full Hamiltonian inversion method, including the detection of the middle region via $v_\mathrm{th}(x)$ and the direct potential inference via $v_\mathrm{cut}(x)$.

\begin{figure}[tphb]
    \centering
    \includegraphics[width=\linewidth]{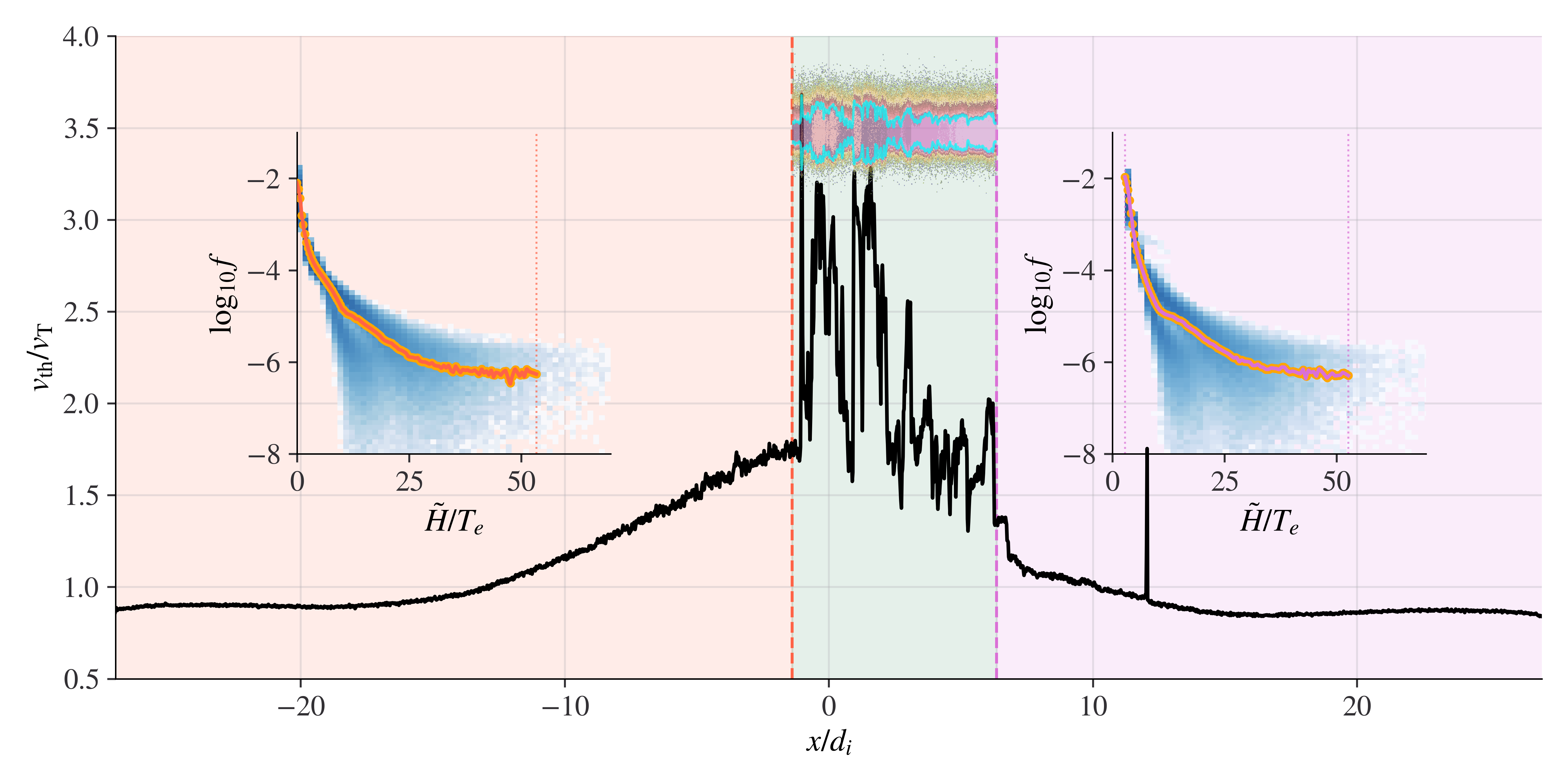}
    \caption{Domain decomposition at simulation time $t\,\omega_{pi} = 10000$, corresponding to $r/R_l = 5.5$ downstream of the Moon. The main panel shows the $v_\mathrm{th}(x)$ profile (black curve), which exhibits sharp enhancements at the locations of the ion acoustic shocks. The middle region ($x_L/d_i = -1.4$ to $x_R/d_i = 6.4$, green shading) is detected as the spatially connected region of elevated and sharply varying $v_\mathrm{th}(x)$ near the potential minimum, flanked by the left (red shading) and right (purple shading) domains. The left and right insets show the distribution of phase space pixels in the $(\tilde{H}, f)$ plane for each domain, with the color map indicating pixel counts per bin and the solid curve showing $f_\mathrm{interp}(\tilde{H})$. The middle inset shows the electron phase space density $f(x, p_x)$ within the middle region, with the cyan curves marking the detected cutoff velocity $\pm v_\mathrm{cut}(x)$ that delineates the flat-top trapped population from the surrounding passing electrons.}
    \label{fig:vth-t10000}
\end{figure}

Figure~\ref{fig:vth-t10000} shows the $v_\mathrm{th}(x)$ profile and the detected domain boundaries. Our algorithm identifies the interval $-1.4\,d_i < x < 6.4\,d_i$ as the middle region, where substantial electron heating occurs between the two shocks, with sharp gradients in $v_\mathrm{th}(x)$ at the boundaries $x_L = -1.4\,d_i$ and $x_R = 6.4\,d_i$. The left and right domains are inverted independently using the $f_\mathrm{interp}$-based optimization. Within the middle region, the cutoff velocity $v_\mathrm{cut}(x)$ is identified as the velocity at which $f(x, p_x)$ drops below $\alpha = 0.5$ of its local plateau value $f_0(x)$ (cyan curves in the middle inset of Figure~\ref{fig:vth-t10000}), and the potential is inferred directly via Equation~\eqref{eq:phi_middle}.

\begin{figure}[tphb]
    \centering
    \includegraphics[width=\linewidth]{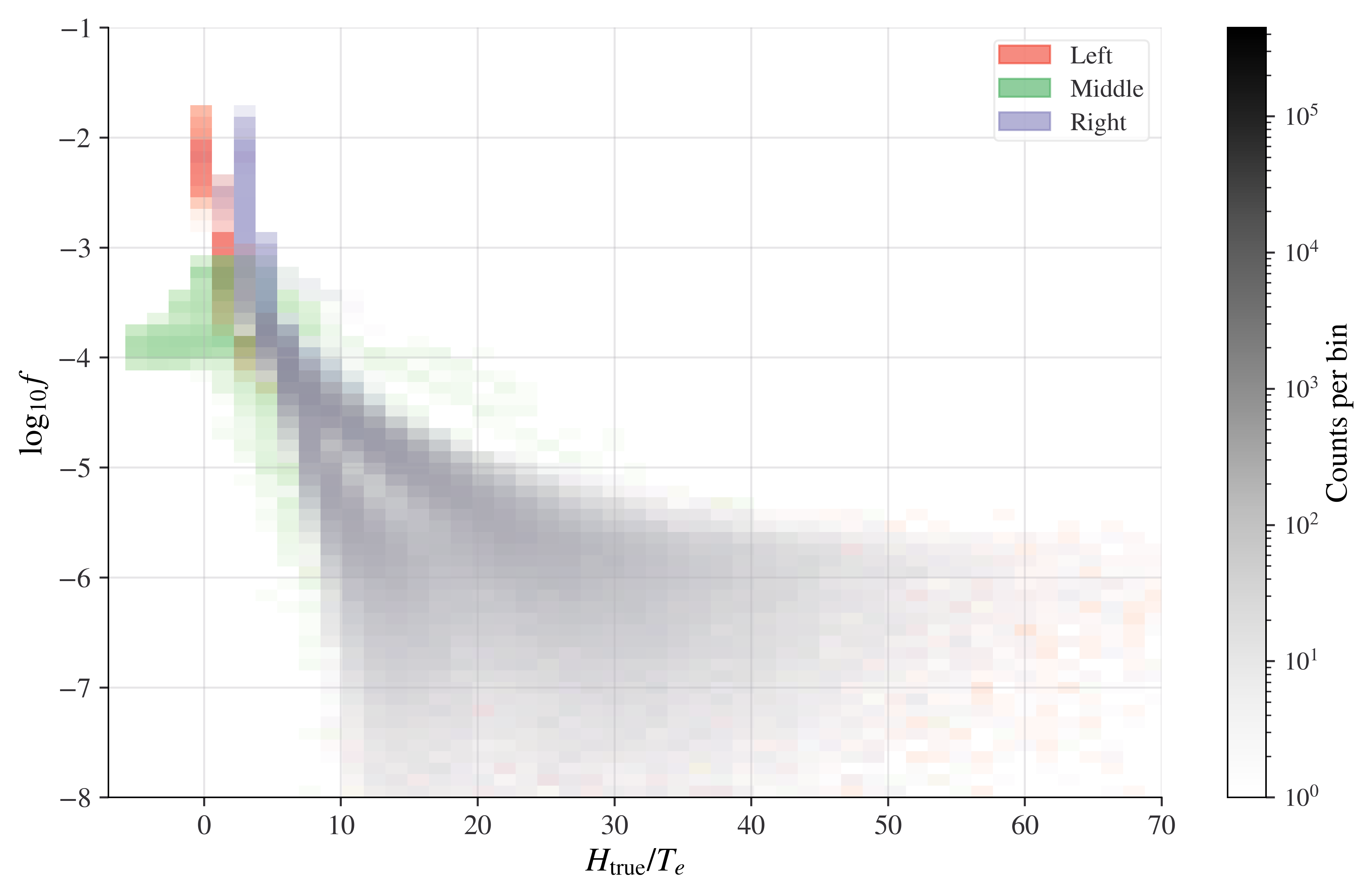}
    \caption{Distribution of electron phase space density $f$ as a function of the true Hamiltonian $H = p_x^2/(2m_e) - e\varphi_\mathrm{true}(x)$ at simulation time $t\,\omega_{pi} = 10000$, corresponding to $r/R_l = 5.5$ downstream of the Moon. The color map indicates the number of phase space pixels per $(H, \log_{10} f)$ bin, with left (red), middle (green), and right (purple) domains shown separately. The left and right domains each trace a well-defined $f(H)$ relation, justifying the independent inversions. The middle domain (green) exhibits a broad horizontal distribution at low $H$, characteristic of the flat-top trapped electron population, for which $f(H)$ is ill-defined and the $f_\mathrm{interp}$-based optimization is replaced by the $v_\mathrm{cut}$-based direct inversion.}
    \label{fig:f-vs-Htrue-t10000}
\end{figure}

Figure~\ref{fig:f-vs-Htrue-t10000} further illustrates why the domain decomposition is essential at this stage. When the true Hamiltonian $H = p_x^2/(2m_e) - e\varphi_\mathrm{true}(x)$ is used, the left and right domains each trace a well-defined $f(H)$ relation, confirming that the quasi-equilibrium assumption $f = f(H)$ holds separately on each side. In contrast, the middle domain exhibits a broad horizontal distribution at low $H$, characteristic of the flat-top trapped electron population for which $f(H)$ is ill-defined, directly justifying the switch to the $v_\mathrm{cut}$-based inversion in this region.

Figure~\ref{fig:modeldata-t10000}(b) compares the inferred potential with the ground truth. The left and right domains exhibit less strahl-driven asymmetry than in Case~1, and both the Hamiltonian inversion and shift methods recover the large-scale potential structure reasonably well on these sides. The primary challenge lies in the middle region, where the potential enhancements associated with the electrostatic shocks must be inferred from $v_\mathrm{cut}(x)$ alone. The Hamiltonian inversion method captures these shock-associated potential enhancements reasonably well, with residual errors arising primarily from uncertainty in the flat-top edge detection near the shock boundaries.

The composite wake potential inferred from the Hamiltonian shift method is consistent with the Hamiltonian inversion method [Figure~\ref{fig:modeldata-t10000}(b)], and notably captures the potential enhancement across shocks in the central region.
%\textcolor{red}{[Todo (Shaosui): assessment of Hamiltonian shift method performance in the middle region.]}

\section{Application to ARTEMIS data}\label{sec:examples}
% A section on what new science is likely to accrue. You must include “at least one illustrative example,” to quote from the paper type description website above. This section closes the gap between the earlier two “must have” sections. That is, given the the current state of scientific discovery in the relevant subdiscipline of space physics and the cutting edge aspects of this new technique or data set, you must then discuss how this new technique will eventually lead to better scientific understanding.
We apply the Hamiltonian inversion method to electron measurements from two ARTEMIS lunar wake crossings \cite{liu2025artemis}, chosen to correspond to the same two evolutionary stages examined in the method validation of Section~\ref{sec:validate}. The first case captures the early stage of plasma refilling, with a minimum central wake density of $\sim 0.001\,n_\mathrm{sw}$, where $n_\mathrm{sw}$ is the solar wind density. The second case captures a later stage, with a minimum central wake density of $\sim 0.01\,n_\mathrm{sw}$. In both cases, the interplanetary magnetic field (IMF) is dominated by a stable $B_y$ component in Selenocentric Solar Ecliptic (SSE) coordinates, ensuring that field-aligned parallel refilling is the dominant process in the orbital plane of the ARTEMIS spacecraft.

We state explicitly the assumptions involved in applying the Hamiltonian inversion method to space-based electron measurements. We assume that solar wind conditions are approximately steady over the duration of the wake crossing, so that the electron phase space density at a given distance downstream of the Moon does not change significantly during the traversal. This assumption is validated by the availability of two ARTEMIS spacecraft \cite{liu2025artemis}: while one spacecraft (P1 or P2) traverses the wake, the other remains in the solar wind and continuously monitors the upstream conditions, allowing us to select events during which the solar wind is sufficiently steady. Under this steady-state assumption, the sequence of electron distribution functions $f(p_\parallel)$ measured along the spacecraft trajectory can be interpreted as a spatial sequence at a single instant.

\begin{figure}[tphb]
    \centering
    \includegraphics[width=\linewidth]{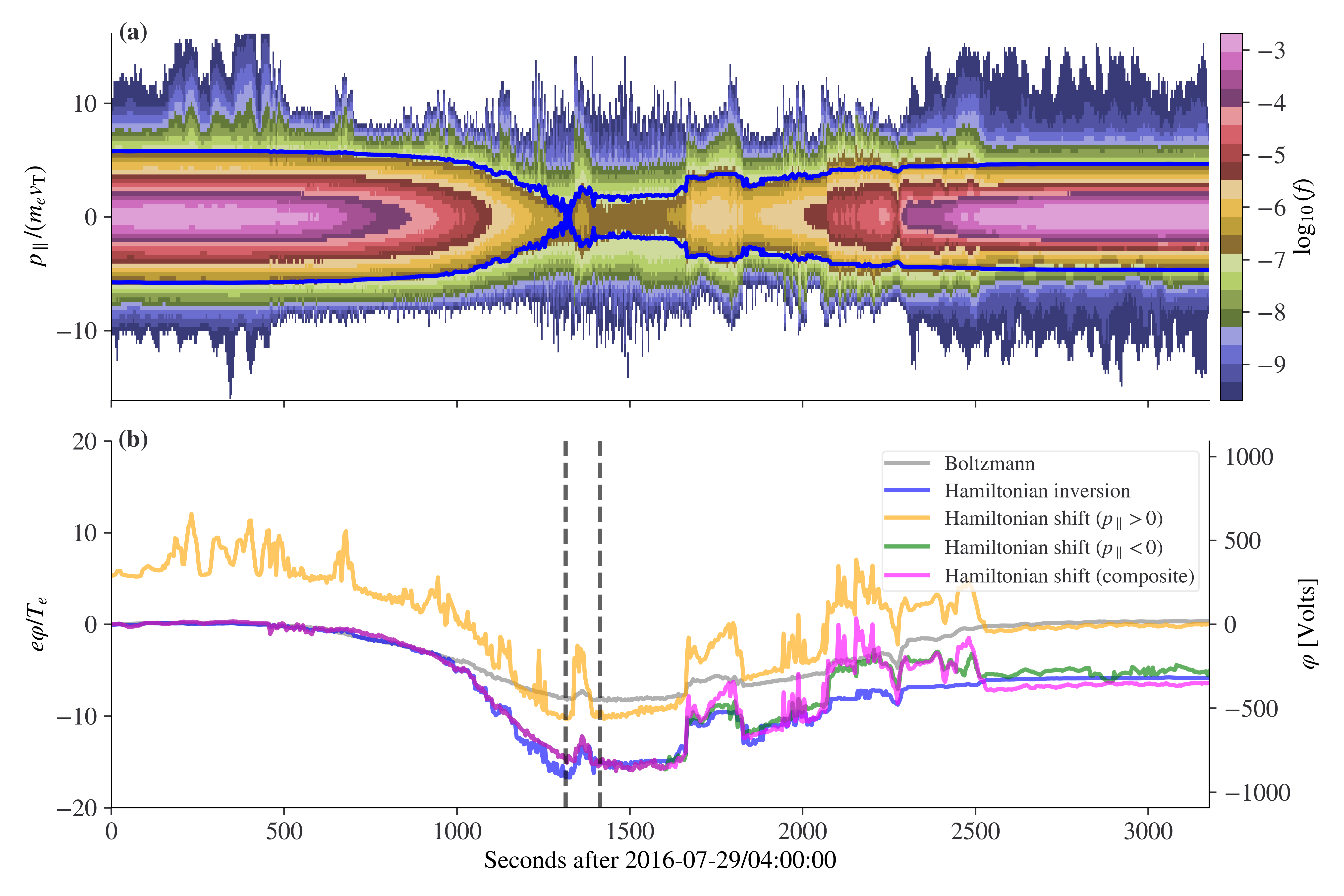}
    \caption{(a) Electron phase space density $f(x, p_\parallel)$ measured by ARTEMIS P1 during the early stage of lunar wake plasma refilling. The horizontal axis is the spacecraft time along the wake traversal, translated to the field-aligned spatial coordinate $x$. A shock structure is identified in the central wake. The blue curve marks the separatrix, defined as the locus of points with $p_\parallel = 0$ at the location of the global potential minimum $\tilde{\varphi}_\mathrm{min}$. (b) Electric potential $\tilde{\varphi}(x)$ inferred by the Boltzmann relation (gray), the Hamiltonian inversion method (blue), and the Hamiltonian shift method using anti-parallel ($p_\parallel < 0$, green) and parallel ($p_\parallel > 0$, yellow) electrons. Vertical dashed lines mark the detected middle region boundaries $x_L$ and $x_R$. The stitching point for the Hamiltonian shift method is $1500$\,s after 2016-07-29/04:00:00.}
    \label{fig:artemmis-ps-pot-early}
\end{figure}

\begin{figure}
    \centering
    \includegraphics[width=\linewidth]{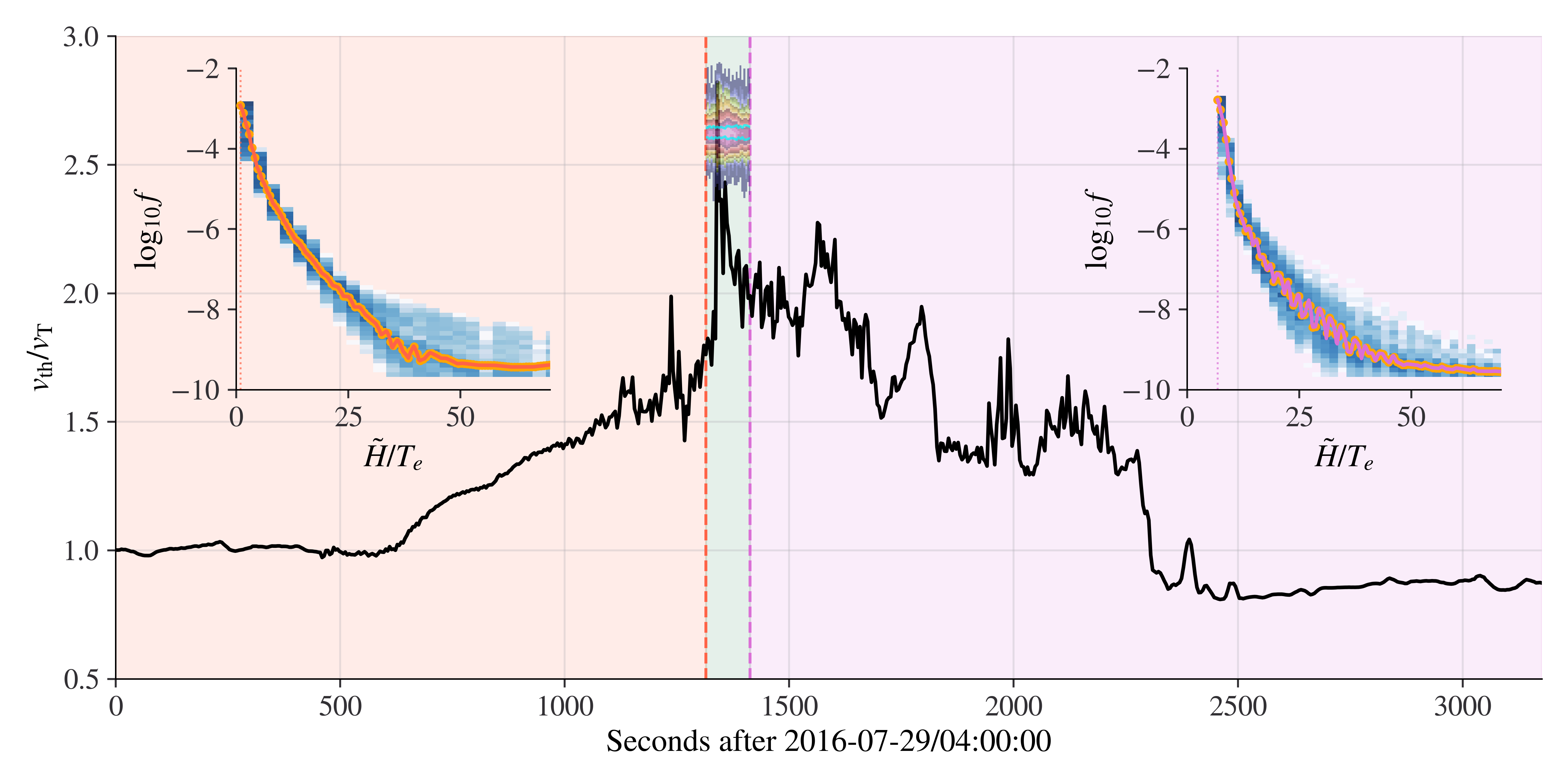}
    \caption{Domain decomposition for the early-stage ARTEMIS wake crossing. The main panel shows the $v_\mathrm{th}(x)$ profile (black curve), with the detected middle region (green shading) flanked by the left (red shading) and right (purple shading) domains. The boundaries $x_L$ and $x_R$ are identified from the sharp enhancement and steep gradient of $v_\mathrm{th}(x)$ associated with the central wake shock. The left and right insets show the distribution of phase space pixels in the $(\tilde{H}, f)$ plane for each domain, with the color map indicating pixel counts per bin and the solid curve showing $f_\mathrm{interp}(\tilde{H})$. The middle inset shows the electron phase space density $f(x, p_\parallel)$ within the middle region, with the cyan curves marking the detected cutoff velocity $\pm v_\mathrm{cut}(x)$ that delineates the flat-top trapped population from the surrounding passing electrons.}
    \label{fig:artemmis-vth-early}
\end{figure}

Figure~\ref{fig:artemmis-ps-pot-early}(a) shows the electron phase space density measured by ARTEMIS P1 during the early stage of plasma refilling. A shock is identified in the central wake [Figure~\ref{fig:artemmis-vth-early} and Event~1 in \citeA{liu2025artemis}], and the domain is accordingly decomposed into left, middle, and right regions. The $f_\mathrm{interp}$-based optimization is applied independently to the left and right domains, and the potential in the middle region is inferred directly from $v_\mathrm{cut}$ via Equation~\eqref{eq:phi_middle}. Figure~\ref{fig:artemmis-ps-pot-early}(b) shows the resulting inferred potential profile. The normalized potential drop from the solar wind to the central wake is $e\Delta\varphi/T_e \sim 15$, corresponding to $\Delta\varphi \sim 800$\,V in absolute units, consistent with the simulation result at the same evolutionary stage [Figure~\ref{fig:modeldata-t3000}(b)]. The asymmetry in the electron distributions between the two sides of the wake produces a corresponding asymmetry in the inferred potential, which is captured by the Hamiltonian inversion method. The inferred potential from the Hamiltonian inversion method agrees well with that from the Hamiltonian shift method over most of the traversal. Discrepancies arise in the interval between $2000$ and $2500$\,s, where the wake appears to be magnetically connected to the terrestrial foreshock. Backstreaming ions from the Earth's bow shock likely perturb the local plasma conditions through Type~II entry \cite{nishino2009solar,halekas2015moon} during this interval, violating the steady-state assumption underlying both methods.

\begin{figure}[tphb]
    \centering
    \includegraphics[width=\linewidth]{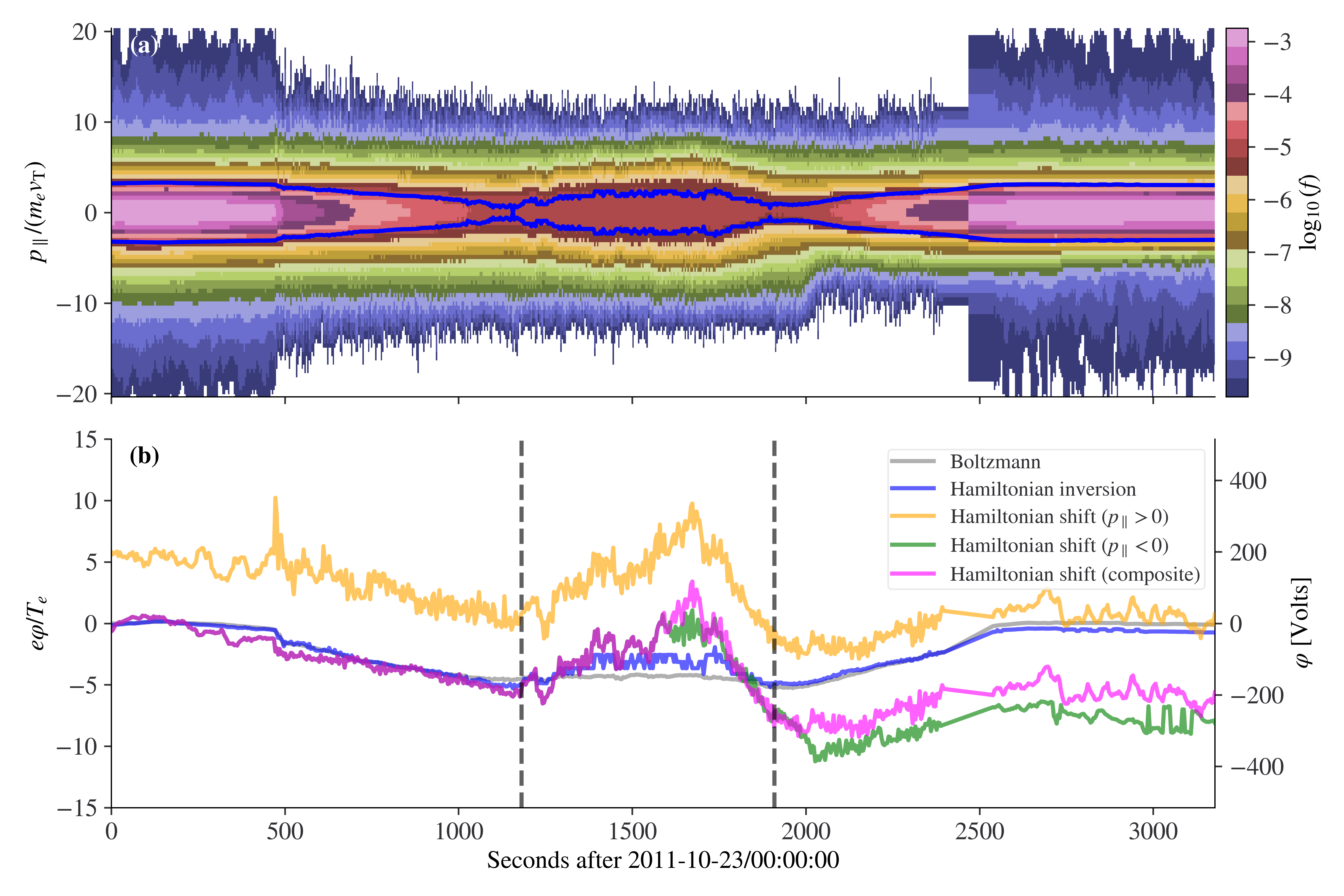}
    \caption{(a) Electron phase space density $f(x, p_\parallel)$ measured by ARTEMIS P1 during the later stage of lunar wake plasma refilling. Ion acoustic shocks have developed in the central wake and flat-top electron distributions are clearly visible in the middle region. The blue curve marks the separatrix, defined as the locus of points with $p_\parallel = 0$ at the location of the global potential minimum $\tilde{\varphi}_\mathrm{min}$. (b) Electric potential $\tilde{\varphi}(x)$ inferred by the Boltzmann relation (gray), the Hamiltonian inversion method (blue), and the Hamiltonian shift method using anti-parallel ($p_\parallel < 0$, green) and parallel ($p_\parallel > 0$, yellow) electrons. Vertical dashed lines mark the detected middle region boundaries $x_L$ and $x_R$. The stitching point for the Hamiltonian shift method is $1500$\,s after 2011-10-23/00:00:00.}
    \label{fig:artemmis-ps-pot-late}
\end{figure}

\begin{figure}[tphb]
    \centering
    \includegraphics[width=\linewidth]{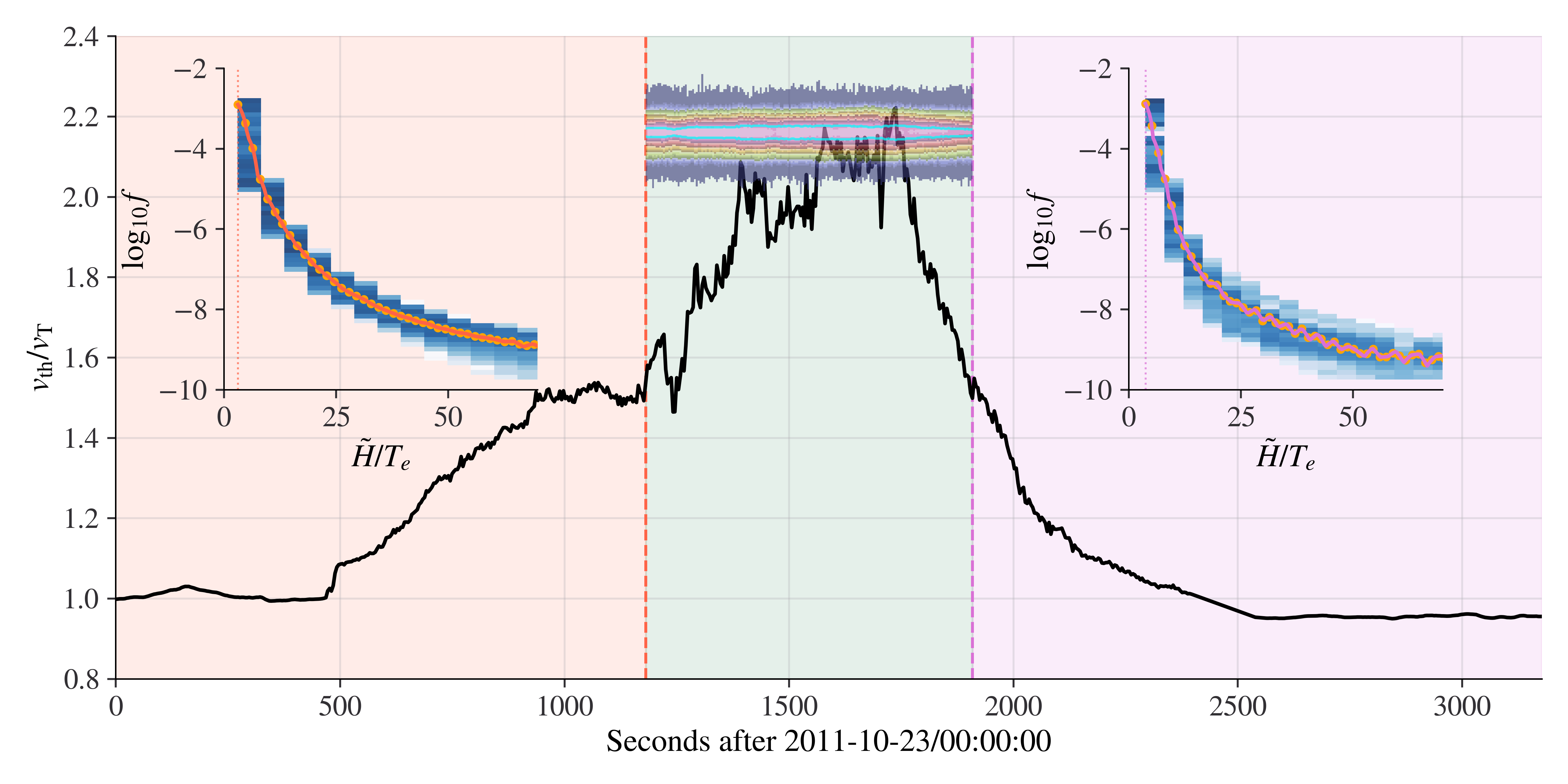}
    \caption{Domain decomposition for the later-stage ARTEMIS wake crossing. The main panel shows the $v_\mathrm{th}(x)$ profile (black curve), which exhibits sharp enhancements at the locations of the ion acoustic shocks. The middle region (green shading) is detected as the spatially connected region of elevated and sharply varying $v_\mathrm{th}(x)$ near the potential minimum, flanked by the left (red shading) and right (purple shading) domains. The left and right insets show the distribution of phase space pixels in the $(\tilde{H}, f)$ plane for each domain, with the color map indicating pixel counts per bin and the solid curve showing $f_\mathrm{interp}(\tilde{H})$. The middle inset shows the electron phase space density $f(x, p_\parallel)$ within the middle region, with the cyan curves marking the detected cutoff velocity $\pm v_\mathrm{cut}(x)$ that delineates the flat-top trapped population from the surrounding passing electrons.}
    \label{fig:artemmis-vth-late}
\end{figure}

Figure~\ref{fig:artemmis-ps-pot-late}(a) shows the electron phase space density measured by ARTEMIS P1 during the later stage of plasma refilling. The domain decomposition follows the same procedure as Case~1: the middle region with flat-top distributions is identified from the $v_\mathrm{th}$ profile [Figure~\ref{fig:artemmis-vth-late}], the potential there is inferred directly from $v_\mathrm{cut}$ via Equation~\eqref{eq:phi_middle}, and the $f_\mathrm{interp}$-based optimization is applied independently to the left and right domains. Figure~\ref{fig:artemmis-ps-pot-late}(b) shows the resulting inferred potential profile, which exhibits three notable features. First, the normalized potential drop from the solar wind to the central wake is $e\Delta\varphi/T_e \sim 5$, corresponding to $\Delta\varphi \sim 200$\,V in absolute units, smaller than in Case~1 and consistent with the simulation result at the same evolutionary stage [Figure~\ref{fig:modeldata-t10000}(b)]. Second, the potential enhancements associated with shock compression in the central wake are captured by the $v_\mathrm{cut}$-based inversion. Third, the asymmetry in the inferred potential between the two sides of the wake is much less pronounced than in Case~1, again consistent with the simulation [Figure~\ref{fig:modeldata-t10000}(b)], reflecting the reduced strahl asymmetry at this later stage. The inferred potentials from the Hamiltonian inversion and Hamiltonian shift methods agree well in the left and right domains up to a constant offset. The most significant discrepancy occurs near the boundaries of the middle region, where the potential gradient inferred from the variation of $v_\mathrm{cut}(x)$ is smaller than that inferred from the energy shift of high-energy electron phase space density contours by the Hamiltonian shift method. This discrepancy likely reflects the difficulty of precisely locating the flat-top edge near the shock boundaries, where the transition from the trapped to the passing population is gradual rather than sharp.

\section{Summary and Conclusions}\label{sec:conclusion}
We have developed the Hamiltonian inversion method for inferring the electric potential structure of the lunar wake from electron phase space density measurements. The method is based on the quasi-static Vlasov equilibrium condition $f = f(H)$, which constrains the electron phase space density to depend only on the local Hamiltonian $H = p_x^2/(2m_e) - e\varphi(x)$ when electrons are in quasi-equilibrium with the electric potential. Rather than comparing the local electron distribution against a fixed reference distribution measured outside the wake as in the Hamiltonian shift method \cite{halekas2014first,xu2019mapping}, the Hamiltonian inversion method solves for the full spatial potential profile $\varphi(x)$ through a domain-decomposition strategy that adapts the inversion technique to the local plasma physics in each region. First, the solar wind strahl renders $f(H)$ multi-valued across the full domain, which is resolved by decomposing the domain at the potential minimum $x^*$ and inverting the left and right domains independently, each with its own $f_\mathrm{interp}(\tilde{H})$ optimized to minimize the misfit between the observed and predicted phase space densities. Second, when counter-streaming ion beams converge near the wake center and generate ion acoustic shocks, electrons are trapped through nonlinear Landau resonance and produce flat-top distributions for which $f(H)$ is ill-defined; the middle region is detected from the sharp, localized enhancement of the effective electron thermal velocity $v_\mathrm{th}(x)$, and the potential there is inferred directly from the cutoff velocity $v_\mathrm{cut}(x)$ [Equation~\eqref{eq:phi_middle}], which encodes the local potential relative to the potential minimum through the width of the flat-top distribution.

We validated the method against PIC simulation data in two cases representing distinct evolutionary stages of the lunar wake. In the early stage ($t\,\omega_{pi} = 3000$, $r/R_l = 1.65$), where no shocks have formed but strong strahl asymmetry is present, the Hamiltonian inversion method accurately captures the asymmetric potential profile on the two sides of the wake, outperforming the Hamiltonian shift method which underestimates the potential gradient in the wake center. In the late stage ($t\,\omega_{pi} = 10000$, $r/R_l = 5.5$), where ion acoustic shocks have developed and flat-top distributions are present in the central wake, the full domain-decomposition algorithm recovers the shock-associated potential enhancements reasonably well, with residual errors arising primarily from uncertainty in the flat-top edge detection near the shock boundaries.

We then applied the method to electron measurements from two ARTEMIS lunar wake crossings corresponding to two similar evolutionary stages as in the simulation. In the early-stage crossing, the inferred potential drop from the solar wind to the central wake is $e\Delta\varphi/T_e \sim 15$ ($\Delta\varphi \sim 800$\,V), with a clear asymmetry between the two sides of the wake driven by the strahl. In the later-stage crossing, the potential drop is $e\Delta\varphi/T_e \sim 5$ ($\Delta\varphi \sim 200$\,V), with less pronounced asymmetry and visible shock-associated potential enhancements in the central wake. In both cases the Hamiltonian inversion method agrees well with the Hamiltonian shift method over most of the traversal, with discrepancies arising near magnetic discontinuities and at the boundaries of the middle region where the flat-top edge detection introduces the largest uncertainty.

The Hamiltonian inversion method can be applied systematically to a larger statistical sample of ARTEMIS wake crossings to characterize the evolution of the lunar wake potential structure as a function of radial distance for different solar wind conditions \cite<see>[for the usage of Hamiltonian shift method for statistical maps of lunar wake electric potentials]{xu2019mapping}. More broadly, the method is applicable to any plasma environment where electrons are in quasi-static equilibrium with a field-aligned electric potential and the phase space density can be measured along a field line, including the wakes of other unmagnetized bodies such as asteroids and comets.

\section*{Open Research Section}
% This section MUST contain a statement that describes where the data supporting the conclusions can be obtained. Data cannot be listed as ''Available from authors'' or stored solely in supporting information. Citations to archived data should be included in your reference list. Wiley will publish it as a separate section on the paper’s page. Examples and complete information are here:
% https://www.agu.org/Publish with AGU/Publish/Author Resources/Data for Authors
The data product and associated Jupyter notebooks used in the analysis are available at Dryad via \citeA{An2026data} [\url{https://tinyurl.com/dryad-lunar-wake-potential}].

\section*{Conflict of Interest declaration}
% Please include a comprehensive conflict of interest statement that reflect all conflicts of interest for all involved authors. If there are no conflicts of interest please state, “The authors declare there are no conflicts of interest for this manuscript.”
The authors declare there are no conflicts of interest for this manuscript.

\acknowledgments
% Enter acknowledgments here. This section is to acknowledge funding, thank colleagues, enter any secondary affiliations, and so on.
This work was supported by NASA contract NAS5-02099 and NASA grant NO.~80NSSC22K1634. Shaosui Xu gratefully acknowledges support from NASA's Lunar Data Analysis Program (LDAP), grant \#80NSSC25K7047. The work by Ferdinand Plaschke was financially supported by the German Center for Aviation and Space (DLR) under contract 50 OC 2201. Data access and processing was done using SPEDAS V4.1 \cite{Angelopoulos19}. We would like to acknowledge high-performance computing support from Derecho (\url{https://doi.org/10.5065/qx9a-pg09}) provided by NCAR's Computational and Information Systems Laboratory, sponsored by the National Science Foundation \cite{derecho}.

%%%%%%%%%%%%%%%%%%%%%%%%%%%%%%%%%%%%%%%%%%%%%%%
% REFERENCES and BIBLIOGRAPHY
%
% \bibliography{<name of your .bib file>} don't specify the file extension
% don't specify bibliographystyle
%
%%%%%%%%%%%%%%%%%%%%%%%%%%%%%%%%%%%%%%%%%%%%%%%

%\bibliography{ enter your bibtex bibliography filename here }
% \bibliography{ref_lunar,full}

\begin{thebibliography}{}
	
	\bibitem [\protect \citeauthoryear {%
		{An}%
		\ \protect \BOthers {.}}{%
		{An}%
		\ \protect \BOthers {.}}{%
		{\protect \APACyear {2025}}%
	}]{%
		An2025plasma}
	\APACinsertmetastar {%
		An2025plasma}%
	\begin{APACrefauthors}%
		{An}, X.%
		, {Angelopoulos}, V.%
		, {Liu}, T\BPBI Z.%
		, {Artemyev}, A.%
		, {Poppe}, A\BPBI R.%
		\BCBL {}\ \BBA {} {Ma}, D.%
	\end{APACrefauthors}%
	\unskip\
	\newblock
	\APACrefYearMonthDay{2025}{{\APACmonth{07}}}{}.
	\newblock
	{\BBOQ}\APACrefatitle {{Plasma Refilling of the Lunar Wake: Plasma-Vacuum
			Interactions, Electrostatic Shocks, and Electromagnetic Instabilities}}
	{{Plasma Refilling of the Lunar Wake: Plasma-Vacuum Interactions,
			Electrostatic Shocks, and Electromagnetic Instabilities}}.{\BBCQ}
	\newblock
	\APACjournalVolNumPages{Journal of Geophysical Research (Space
		Physics)}{130}{7}{e2025JA034205}.
	\newblock
	\begin{APACrefDOI} \doi{10.1029/2025JA034205} \end{APACrefDOI}
	\PrintBackRefs{\CurrentBib}
	
	\bibitem [\protect \citeauthoryear {%
		An%
		\ \protect \BOthers {.}}{%
		An%
		\ \protect \BOthers {.}}{%
		{\protect \APACyear {2026}}%
	}]{%
		An2026data}
	\APACinsertmetastar {%
		An2026data}%
	\begin{APACrefauthors}%
		An, X.%
		, Xu, S.%
		, Angelopoulos, V.%
		, Liu, T\BPBI Z\BPBI L.%
		, Poppe, A\BPBI R.%
		, Halekas, J\BPBI S.%
		\BCBL {}\ \BBA {} Plaschke, F.%
	\end{APACrefauthors}%
	\unskip\
	\newblock
	\APACrefYearMonthDay{2026}{}{}.
	\newblock
	\APACrefbtitle {Data and code from: Inferring lunar wake potentials from
		electron phase space densities.} {Data and code from: Inferring lunar wake
		potentials from electron phase space densities.}
	\newblock
	\APACaddressPublisher{}{Dryad Digital Repository}.
	\newblock
	\begin{APACrefURL} \url{https://tinyurl.com/dryad-lunar-wake-potential}
	\end{APACrefURL}
	\PrintBackRefs{\CurrentBib}
	
	\bibitem [\protect \citeauthoryear {%
		{Angelopoulos}%
		\ \protect \BOthers {.}}{%
		{Angelopoulos}%
		\ \protect \BOthers {.}}{%
		{\protect \APACyear {2019}}%
	}]{%
		Angelopoulos19}
	\APACinsertmetastar {%
		Angelopoulos19}%
	\begin{APACrefauthors}%
		{Angelopoulos}, V.%
		, {Cruce}, P.%
		, {Drozdov}, A.%
		, {Grimes}, E\BPBI W.%
		, {Hatzigeorgiu}, N.%
		, {King}, D\BPBI A.%
		\BDBL {}{Schroeder}, P.%
	\end{APACrefauthors}%
	\unskip\
	\newblock
	\APACrefYearMonthDay{2019}{{\APACmonth{01}}}{}.
	\newblock
	{\BBOQ}\APACrefatitle {{The Space Physics Environment Data Analysis System
			(SPEDAS)}} {{The Space Physics Environment Data Analysis System
			(SPEDAS)}}.{\BBCQ}
	\newblock
	\APACjournalVolNumPages{\ssr}{215}{}{9}.
	\newblock
	\begin{APACrefDOI} \doi{10.1007/s11214-018-0576-4} \end{APACrefDOI}
	\PrintBackRefs{\CurrentBib}
	
	\bibitem [\protect \citeauthoryear {%
		Birch%
		\ \BBA {} Chapman%
	}{%
		Birch%
		\ \BBA {} Chapman%
	}{%
		{\protect \APACyear {2001}}%
		{\protect \APACexlab {{\protect \BCnt {1}}}}}]{%
		birch2001detailed}
	\APACinsertmetastar {%
		birch2001detailed}%
	\begin{APACrefauthors}%
		Birch, P\BPBI C.%
		\BCBT {}\ \BBA {} Chapman, S\BPBI C.%
	\end{APACrefauthors}%
	\unskip\
	\newblock
	\APACrefYearMonthDay{2001{\protect \BCnt {1}}}{}{}.
	\newblock
	{\BBOQ}\APACrefatitle {Detailed structure and dynamics in particle-in-cell
		simulations of the lunar wake} {Detailed structure and dynamics in
		particle-in-cell simulations of the lunar wake}.{\BBCQ}
	\newblock
	\APACjournalVolNumPages{Physics of Plasmas}{8}{10}{4551--4559}.
	\PrintBackRefs{\CurrentBib}
	
	\bibitem [\protect \citeauthoryear {%
		Birch%
		\ \BBA {} Chapman%
	}{%
		Birch%
		\ \BBA {} Chapman%
	}{%
		{\protect \APACyear {2001}}%
		{\protect \APACexlab {{\protect \BCnt {2}}}}}]{%
		birch2001particle}
	\APACinsertmetastar {%
		birch2001particle}%
	\begin{APACrefauthors}%
		Birch, P\BPBI C.%
		\BCBT {}\ \BBA {} Chapman, S\BPBI C.%
	\end{APACrefauthors}%
	\unskip\
	\newblock
	\APACrefYearMonthDay{2001{\protect \BCnt {2}}}{}{}.
	\newblock
	{\BBOQ}\APACrefatitle {Particle-in-cell simulations of the lunar wake with high
		phase space resolution} {Particle-in-cell simulations of the lunar wake with
		high phase space resolution}.{\BBCQ}
	\newblock
	\APACjournalVolNumPages{Geophysical research letters}{28}{2}{219--222}.
	\PrintBackRefs{\CurrentBib}
	
	\bibitem [\protect \citeauthoryear {%
		Bird%
		\ \protect \BOthers {.}}{%
		Bird%
		\ \protect \BOthers {.}}{%
		{\protect \APACyear {2021}}%
	}]{%
		bird2021vpic}
	\APACinsertmetastar {%
		bird2021vpic}%
	\begin{APACrefauthors}%
		Bird, R.%
		, Tan, N.%
		, Luedtke, S\BPBI V.%
		, Harrell, S\BPBI L.%
		, Taufer, M.%
		\BCBL {}\ \BBA {} Albright, B.%
	\end{APACrefauthors}%
	\unskip\
	\newblock
	\APACrefYearMonthDay{2021}{}{}.
	\newblock
	{\BBOQ}\APACrefatitle {VPIC 2.0: Next generation particle-in-cell simulations}
	{Vpic 2.0: Next generation particle-in-cell simulations}.{\BBCQ}
	\newblock
	\APACjournalVolNumPages{IEEE Transactions on Parallel and Distributed
		Systems}{33}{4}{952--963}.
	\PrintBackRefs{\CurrentBib}
	
	\bibitem [\protect \citeauthoryear {%
		{Bonnell}%
		\ \protect \BOthers {.}}{%
		{Bonnell}%
		\ \protect \BOthers {.}}{%
		{\protect \APACyear {2008}}%
	}]{%
		Bonnell08}
	\APACinsertmetastar {%
		Bonnell08}%
	\begin{APACrefauthors}%
		{Bonnell}, J\BPBI W.%
		, {Mozer}, F\BPBI S.%
		, {Delory}, G\BPBI T.%
		, {Hull}, A\BPBI J.%
		, {Ergun}, R\BPBI E.%
		, {Cully}, C\BPBI M.%
		\BDBL {}{Harvey}, P\BPBI R.%
	\end{APACrefauthors}%
	\unskip\
	\newblock
	\APACrefYearMonthDay{2008}{{\APACmonth{12}}}{}.
	\newblock
	{\BBOQ}\APACrefatitle {{The Electric Field Instrument (EFI) for THEMIS}} {{The
			Electric Field Instrument (EFI) for THEMIS}}.{\BBCQ}
	\newblock
	\APACjournalVolNumPages{\ssr}{141}{}{303-341}.
	\newblock
	\begin{APACrefDOI} \doi{10.1007/s11214-008-9469-2} \end{APACrefDOI}
	\PrintBackRefs{\CurrentBib}
	
	\bibitem [\protect \citeauthoryear {%
		Bowers%
		, Albright%
		, Bergen%
		\BCBL {}\ \protect \BOthers {.}}{%
		Bowers%
		, Albright%
		, Bergen%
		\BCBL {}\ \protect \BOthers {.}}{%
		{\protect \APACyear {2008}}%
	}]{%
		bowers20080}
	\APACinsertmetastar {%
		bowers20080}%
	\begin{APACrefauthors}%
		Bowers, K\BPBI J.%
		, Albright, B\BPBI J.%
		, Bergen, B.%
		, Yin, L.%
		, Barker, K\BPBI J.%
		\BCBL {}\ \BBA {} Kerbyson, D\BPBI J.%
	\end{APACrefauthors}%
	\unskip\
	\newblock
	\APACrefYearMonthDay{2008}{}{}.
	\newblock
	{\BBOQ}\APACrefatitle {0.374 pflop/s trillion-particle kinetic modeling of
		laser plasma interaction on roadrunner} {0.374 pflop/s trillion-particle
		kinetic modeling of laser plasma interaction on roadrunner}.{\BBCQ}
	\newblock
	\BIn{} \APACrefbtitle {SC'08: Proceedings of the 2008 ACM/IEEE Conference on
		Supercomputing} {Sc'08: Proceedings of the 2008 acm/ieee conference on
		supercomputing}\ (\BPGS\ 1--11).
	\PrintBackRefs{\CurrentBib}
	
	\bibitem [\protect \citeauthoryear {%
		Bowers%
		, Albright%
		, Yin%
		, Bergen%
		\BCBL {}\ \BBA {} Kwan%
	}{%
		Bowers%
		, Albright%
		, Yin%
		\BCBL {}\ \protect \BOthers {.}}{%
		{\protect \APACyear {2008}}%
	}]{%
		bowers2008ultrahigh}
	\APACinsertmetastar {%
		bowers2008ultrahigh}%
	\begin{APACrefauthors}%
		Bowers, K\BPBI J.%
		, Albright, B\BPBI J.%
		, Yin, L.%
		, Bergen, B.%
		\BCBL {}\ \BBA {} Kwan, T\BPBI J.%
	\end{APACrefauthors}%
	\unskip\
	\newblock
	\APACrefYearMonthDay{2008}{}{}.
	\newblock
	{\BBOQ}\APACrefatitle {Ultrahigh performance three-dimensional electromagnetic
		relativistic kinetic plasma simulation} {Ultrahigh performance
		three-dimensional electromagnetic relativistic kinetic plasma
		simulation}.{\BBCQ}
	\newblock
	\APACjournalVolNumPages{Physics of Plasmas}{15}{5}{}.
	\PrintBackRefs{\CurrentBib}
	
	\bibitem [\protect \citeauthoryear {%
		Bowers%
		\ \protect \BOthers {.}}{%
		Bowers%
		\ \protect \BOthers {.}}{%
		{\protect \APACyear {2009}}%
	}]{%
		bowers2009advances}
	\APACinsertmetastar {%
		bowers2009advances}%
	\begin{APACrefauthors}%
		Bowers, K\BPBI J.%
		, Albright, B\BPBI J.%
		, Yin, L.%
		, Daughton, W.%
		, Roytershteyn, V.%
		, Bergen, B.%
		\BCBL {}\ \BBA {} Kwan, T\BPBI J.%
	\end{APACrefauthors}%
	\unskip\
	\newblock
	\APACrefYearMonthDay{2009}{}{}.
	\newblock
	{\BBOQ}\APACrefatitle {Advances in petascale kinetic plasma simulation with
		VPIC and Roadrunner} {Advances in petascale kinetic plasma simulation with
		vpic and roadrunner}.{\BBCQ}
	\newblock
	\BIn{} \APACrefbtitle {Journal of Physics: Conference Series} {Journal of
		physics: Conference series}\ (\BVOL~180, \BPG~012055).
	\PrintBackRefs{\CurrentBib}
	
	\bibitem [\protect \citeauthoryear {%
		{Computational and Information Systems Laboratory}%
	}{%
		{Computational and Information Systems Laboratory}%
	}{%
		{\protect \APACyear {2024}}%
	}]{%
		derecho}
	\APACinsertmetastar {%
		derecho}%
	\begin{APACrefauthors}%
		{Computational and Information Systems Laboratory}.%
	\end{APACrefauthors}%
	\unskip\
	\newblock
	\APACrefYearMonthDay{2024}{}{}.
	\newblock
	\APACrefbtitle {Derecho: {HPE} {C}ray {EX} {S}ystem.} {Derecho: {HPE} {C}ray
		{EX} {S}ystem.}
	\newblock
	\APAChowpublished {Boulder, CO: National Center for Atmospheric Research}.
	\newblock
	\begin{APACrefURL} \url{https://doi.org/10.5065/qx9a-pg09} \end{APACrefURL}
	\PrintBackRefs{\CurrentBib}
	
	\bibitem [\protect \citeauthoryear {%
		{Crow}%
		, {Auer}%
		\BCBL {}\ \BBA {} {Allen}%
	}{%
		{Crow}%
		\ \protect \BOthers {.}}{%
		{\protect \APACyear {1975}}%
	}]{%
		crow1975expansion}
	\APACinsertmetastar {%
		crow1975expansion}%
	\begin{APACrefauthors}%
		{Crow}, J\BPBI E.%
		, {Auer}, P\BPBI L.%
		\BCBL {}\ \BBA {} {Allen}, J\BPBI E.%
	\end{APACrefauthors}%
	\unskip\
	\newblock
	\APACrefYearMonthDay{1975}{{\APACmonth{08}}}{}.
	\newblock
	{\BBOQ}\APACrefatitle {{The expansion of a plasma into a vacuum}} {{The
			expansion of a plasma into a vacuum}}.{\BBCQ}
	\newblock
	\APACjournalVolNumPages{Journal of Plasma Physics}{14}{1}{65-76}.
	\newblock
	\begin{APACrefDOI} \doi{10.1017/S0022377800025538} \end{APACrefDOI}
	\PrintBackRefs{\CurrentBib}
	
	\bibitem [\protect \citeauthoryear {%
		{Denavit}%
	}{%
		{Denavit}%
	}{%
		{\protect \APACyear {1979}}%
	}]{%
		denavit1979collisionless}
	\APACinsertmetastar {%
		denavit1979collisionless}%
	\begin{APACrefauthors}%
		{Denavit}, J.%
	\end{APACrefauthors}%
	\unskip\
	\newblock
	\APACrefYearMonthDay{1979}{{\APACmonth{07}}}{}.
	\newblock
	{\BBOQ}\APACrefatitle {{Collisionless plasma expansion into a vacuum}}
	{{Collisionless plasma expansion into a vacuum}}.{\BBCQ}
	\newblock
	\APACjournalVolNumPages{Physics of Fluids}{22}{7}{1384-1392}.
	\newblock
	\begin{APACrefDOI} \doi{10.1063/1.862751} \end{APACrefDOI}
	\PrintBackRefs{\CurrentBib}
	
	\bibitem [\protect \citeauthoryear {%
		Farrell%
		, Kaiser%
		, Steinberg%
		\BCBL {}\ \BBA {} Bale%
	}{%
		Farrell%
		\ \protect \BOthers {.}}{%
		{\protect \APACyear {1998}}%
	}]{%
		farrell1998simple}
	\APACinsertmetastar {%
		farrell1998simple}%
	\begin{APACrefauthors}%
		Farrell, W.%
		, Kaiser, M.%
		, Steinberg, J.%
		\BCBL {}\ \BBA {} Bale, S.%
	\end{APACrefauthors}%
	\unskip\
	\newblock
	\APACrefYearMonthDay{1998}{}{}.
	\newblock
	{\BBOQ}\APACrefatitle {A simple simulation of a plasma void: Applications to
		Wind observations of the lunar wake} {A simple simulation of a plasma void:
		Applications to wind observations of the lunar wake}.{\BBCQ}
	\newblock
	\APACjournalVolNumPages{Journal of Geophysical Research: Space
		Physics}{103}{A10}{23653--23660}.
	\PrintBackRefs{\CurrentBib}
	
	\bibitem [\protect \citeauthoryear {%
		{Gurevich}%
		, {Pari{\v{i}}skaya}%
		\BCBL {}\ \BBA {} {Pitaevski{\v{i}}}%
	}{%
		{Gurevich}%
		\ \protect \BOthers {.}}{%
		{\protect \APACyear {1966}}%
	}]{%
		gurevich1966self}
	\APACinsertmetastar {%
		gurevich1966self}%
	\begin{APACrefauthors}%
		{Gurevich}, A\BPBI V.%
		, {Pari{\v{i}}skaya}, L\BPBI V.%
		\BCBL {}\ \BBA {} {Pitaevski{\v{i}}}, L\BPBI P.%
	\end{APACrefauthors}%
	\unskip\
	\newblock
	\APACrefYearMonthDay{1966}{{\APACmonth{02}}}{}.
	\newblock
	{\BBOQ}\APACrefatitle {{Self-similar Motion of Rarefied Plasma}} {{Self-similar
			Motion of Rarefied Plasma}}.{\BBCQ}
	\newblock
	\APACjournalVolNumPages{Soviet Journal of Experimental and Theoretical
		Physics}{22}{}{449}.
	\PrintBackRefs{\CurrentBib}
	
	\bibitem [\protect \citeauthoryear {%
		Halekas%
		, Brain%
		\BCBL {}\ \BBA {} Holmstr{\"o}m%
	}{%
		Halekas%
		\ \protect \BOthers {.}}{%
		{\protect \APACyear {2015}}%
	}]{%
		halekas2015moon}
	\APACinsertmetastar {%
		halekas2015moon}%
	\begin{APACrefauthors}%
		Halekas, J.%
		, Brain, D.%
		\BCBL {}\ \BBA {} Holmstr{\"o}m, M.%
	\end{APACrefauthors}%
	\unskip\
	\newblock
	\APACrefYearMonthDay{2015}{}{}.
	\newblock
	{\BBOQ}\APACrefatitle {Moon's plasma wake} {Moon's plasma wake}.{\BBCQ}
	\newblock
	\APACjournalVolNumPages{Magnetotails in the solar system}{}{}{149--167}.
	\PrintBackRefs{\CurrentBib}
	
	\bibitem [\protect \citeauthoryear {%
		Halekas%
		, Poppe%
		\BCBL {}\ \BBA {} McFadden%
	}{%
		Halekas%
		\ \protect \BOthers {.}}{%
		{\protect \APACyear {2014}}%
	}]{%
		halekas2014effects}
	\APACinsertmetastar {%
		halekas2014effects}%
	\begin{APACrefauthors}%
		Halekas, J.%
		, Poppe, A.%
		\BCBL {}\ \BBA {} McFadden, J.%
	\end{APACrefauthors}%
	\unskip\
	\newblock
	\APACrefYearMonthDay{2014}{}{}.
	\newblock
	{\BBOQ}\APACrefatitle {The effects of solar wind velocity distributions on the
		refilling of the lunar wake: ARTEMIS observations and comparisons to
		one-dimensional theory} {The effects of solar wind velocity distributions on
		the refilling of the lunar wake: Artemis observations and comparisons to
		one-dimensional theory}.{\BBCQ}
	\newblock
	\APACjournalVolNumPages{Journal of Geophysical Research: Space
		Physics}{119}{7}{5133--5149}.
	\PrintBackRefs{\CurrentBib}
	
	\bibitem [\protect \citeauthoryear {%
		{Halekas}%
		\ \protect \BOthers {.}}{%
		{Halekas}%
		\ \protect \BOthers {.}}{%
		{\protect \APACyear {2011}}%
	}]{%
		halekas2014first}
	\APACinsertmetastar {%
		halekas2014first}%
	\begin{APACrefauthors}%
		{Halekas}, J\BPBI S.%
		, {Angelopoulos}, V.%
		, {Sibeck}, D\BPBI G.%
		, {Khurana}, K\BPBI K.%
		, {Russell}, C\BPBI T.%
		, {Delory}, G\BPBI T.%
		\BDBL {}{Glassmeier}, K\BPBI H.%
	\end{APACrefauthors}%
	\unskip\
	\newblock
	\APACrefYearMonthDay{2011}{{\APACmonth{12}}}{}.
	\newblock
	{\BBOQ}\APACrefatitle {{First Results from ARTEMIS, a New Two-Spacecraft Lunar
			Mission: Counter-Streaming Plasma Populations in the Lunar Wake}} {{First
			Results from ARTEMIS, a New Two-Spacecraft Lunar Mission: Counter-Streaming
			Plasma Populations in the Lunar Wake}}.{\BBCQ}
	\newblock
	\APACjournalVolNumPages{\ssr}{165}{1-4}{93-107}.
	\newblock
	\begin{APACrefDOI} \doi{10.1007/s11214-010-9738-8} \end{APACrefDOI}
	\PrintBackRefs{\CurrentBib}
	
	\bibitem [\protect \citeauthoryear {%
		{Halekas}%
		, {Bale}%
		, {Mitchell}%
		\BCBL {}\ \BBA {} {Lin}%
	}{%
		{Halekas}%
		\ \protect \BOthers {.}}{%
		{\protect \APACyear {2005}}%
	}]{%
		halekas2005electrons}
	\APACinsertmetastar {%
		halekas2005electrons}%
	\begin{APACrefauthors}%
		{Halekas}, J\BPBI S.%
		, {Bale}, S\BPBI D.%
		, {Mitchell}, D\BPBI L.%
		\BCBL {}\ \BBA {} {Lin}, R\BPBI P.%
	\end{APACrefauthors}%
	\unskip\
	\newblock
	\APACrefYearMonthDay{2005}{{\APACmonth{07}}}{}.
	\newblock
	{\BBOQ}\APACrefatitle {{Electrons and magnetic fields in the lunar plasma
			wake}} {{Electrons and magnetic fields in the lunar plasma wake}}.{\BBCQ}
	\newblock
	\APACjournalVolNumPages{Journal of Geophysical Research (Space
		Physics)}{110}{A7}{A07222}.
	\newblock
	\begin{APACrefDOI} \doi{10.1029/2004JA010991} \end{APACrefDOI}
	\PrintBackRefs{\CurrentBib}
	
	\bibitem [\protect \citeauthoryear {%
		Kivelson%
	}{%
		Kivelson%
	}{%
		{\protect \APACyear {2016}}%
	}]{%
		kivelson2016moons}
	\APACinsertmetastar {%
		kivelson2016moons}%
	\begin{APACrefauthors}%
		Kivelson, M\BPBI G.%
	\end{APACrefauthors}%
	\unskip\
	\newblock
	\APACrefYearMonthDay{2016}{}{}.
	\newblock
	{\BBOQ}\APACrefatitle {Moons, asteroids, and comets jnteracting with their
		surroundings} {Moons, asteroids, and comets jnteracting with their
		surroundings}.{\BBCQ}
	\newblock
	\APACjournalVolNumPages{Heliophysics: Active Stars, their Astrospheres, and
		Impacts on Planetary Environments}{}{}{226--250}.
	\PrintBackRefs{\CurrentBib}
	
	\bibitem [\protect \citeauthoryear {%
		Liu%
		, An%
		, Angelopoulos%
		\BCBL {}\ \BBA {} Poppe%
	}{%
		Liu%
		\ \protect \BOthers {.}}{%
		{\protect \APACyear {2025}}%
	}]{%
		liu2025artemis}
	\APACinsertmetastar {%
		liu2025artemis}%
	\begin{APACrefauthors}%
		Liu, T\BPBI Z.%
		, An, X.%
		, Angelopoulos, V.%
		\BCBL {}\ \BBA {} Poppe, A\BPBI R.%
	\end{APACrefauthors}%
	\unskip\
	\newblock
	\APACrefYearMonthDay{2025}{}{}.
	\newblock
	{\BBOQ}\APACrefatitle {ARTEMIS observations of electrostatic shocks inside the
		lunar wake} {Artemis observations of electrostatic shocks inside the lunar
		wake}.{\BBCQ}
	\newblock
	\APACjournalVolNumPages{The Astrophysical Journal Letters}{990}{2}{L36}.
	\PrintBackRefs{\CurrentBib}
	
	\bibitem [\protect \citeauthoryear {%
		Maksimovic%
		, Gary%
		\BCBL {}\ \BBA {} Skoug%
	}{%
		Maksimovic%
		\ \protect \BOthers {.}}{%
		{\protect \APACyear {2000}}%
	}]{%
		maksimovic2000solar}
	\APACinsertmetastar {%
		maksimovic2000solar}%
	\begin{APACrefauthors}%
		Maksimovic, M.%
		, Gary, S\BPBI P.%
		\BCBL {}\ \BBA {} Skoug, R\BPBI M.%
	\end{APACrefauthors}%
	\unskip\
	\newblock
	\APACrefYearMonthDay{2000}{}{}.
	\newblock
	{\BBOQ}\APACrefatitle {Solar wind electron suprathermal strength and
		temperature gradients: Ulysses observations} {Solar wind electron
		suprathermal strength and temperature gradients: Ulysses
		observations}.{\BBCQ}
	\newblock
	\APACjournalVolNumPages{Journal of Geophysical Research: Space
		Physics}{105}{A8}{18337--18350}.
	\PrintBackRefs{\CurrentBib}
	
	\bibitem [\protect \citeauthoryear {%
		Malaspina%
		\ \BBA {} Hutchinson%
	}{%
		Malaspina%
		\ \BBA {} Hutchinson%
	}{%
		{\protect \APACyear {2019}}%
	}]{%
		malaspina2019properties}
	\APACinsertmetastar {%
		malaspina2019properties}%
	\begin{APACrefauthors}%
		Malaspina, D\BPBI M.%
		\BCBT {}\ \BBA {} Hutchinson, I\BPBI H.%
	\end{APACrefauthors}%
	\unskip\
	\newblock
	\APACrefYearMonthDay{2019}{}{}.
	\newblock
	{\BBOQ}\APACrefatitle {Properties of electron phase space holes in the lunar
		plasma environment} {Properties of electron phase space holes in the lunar
		plasma environment}.{\BBCQ}
	\newblock
	\APACjournalVolNumPages{Journal of Geophysical Research: Space
		Physics}{124}{7}{4994--5008}.
	\PrintBackRefs{\CurrentBib}
	
	\bibitem [\protect \citeauthoryear {%
		Mora%
	}{%
		Mora%
	}{%
		{\protect \APACyear {2003}}%
	}]{%
		mora2003plasma}
	\APACinsertmetastar {%
		mora2003plasma}%
	\begin{APACrefauthors}%
		Mora, P.%
	\end{APACrefauthors}%
	\unskip\
	\newblock
	\APACrefYearMonthDay{2003}{May}{}.
	\newblock
	{\BBOQ}\APACrefatitle {Plasma Expansion into a Vacuum} {Plasma expansion into a
		vacuum}.{\BBCQ}
	\newblock
	\APACjournalVolNumPages{Phys. Rev. Lett.}{90}{}{185002}.
	\newblock
	\begin{APACrefURL} \url{https://link.aps.org/doi/10.1103/PhysRevLett.90.185002}
	\end{APACrefURL}
	\newblock
	\begin{APACrefDOI} \doi{10.1103/PhysRevLett.90.185002} \end{APACrefDOI}
	\PrintBackRefs{\CurrentBib}
	
	\bibitem [\protect \citeauthoryear {%
		Nishino%
		\ \protect \BOthers {.}}{%
		Nishino%
		\ \protect \BOthers {.}}{%
		{\protect \APACyear {2009}}%
	}]{%
		nishino2009solar}
	\APACinsertmetastar {%
		nishino2009solar}%
	\begin{APACrefauthors}%
		Nishino, M.%
		, Fujimoto, M.%
		, Maezawa, K.%
		, Saito, Y.%
		, Yokota, S.%
		, Asamura, K.%
		\BDBL {}others%
	\end{APACrefauthors}%
	\unskip\
	\newblock
	\APACrefYearMonthDay{2009}{}{}.
	\newblock
	{\BBOQ}\APACrefatitle {Solar-wind proton access deep into the near-Moon wake}
	{Solar-wind proton access deep into the near-moon wake}.{\BBCQ}
	\newblock
	\APACjournalVolNumPages{Geophysical Research Letters}{36}{16}{}.
	\PrintBackRefs{\CurrentBib}
	
	\bibitem [\protect \citeauthoryear {%
		{\v{S}}tver{\'a}k%
		\ \protect \BOthers {.}}{%
		{\v{S}}tver{\'a}k%
		\ \protect \BOthers {.}}{%
		{\protect \APACyear {2009}}%
	}]{%
		vstverak2009radial}
	\APACinsertmetastar {%
		vstverak2009radial}%
	\begin{APACrefauthors}%
		{\v{S}}tver{\'a}k, {\v{S}}.%
		, Maksimovic, M.%
		, Tr{\'a}vn{\'\i}{\v{c}}ek, P\BPBI M.%
		, Marsch, E.%
		, Fazakerley, A\BPBI N.%
		\BCBL {}\ \BBA {} Scime, E\BPBI E.%
	\end{APACrefauthors}%
	\unskip\
	\newblock
	\APACrefYearMonthDay{2009}{}{}.
	\newblock
	{\BBOQ}\APACrefatitle {Radial evolution of nonthermal electron populations in
		the low-latitude solar wind: Helios, Cluster, and Ulysses observations}
	{Radial evolution of nonthermal electron populations in the low-latitude
		solar wind: Helios, cluster, and ulysses observations}.{\BBCQ}
	\newblock
	\APACjournalVolNumPages{Journal of Geophysical Research: Space
		Physics}{114}{A5}{}.
	\PrintBackRefs{\CurrentBib}
	
	\bibitem [\protect \citeauthoryear {%
		Xu%
		\ \protect \BOthers {.}}{%
		Xu%
		\ \protect \BOthers {.}}{%
		{\protect \APACyear {2019}}%
	}]{%
		xu2019mapping}
	\APACinsertmetastar {%
		xu2019mapping}%
	\begin{APACrefauthors}%
		Xu, S.%
		, Poppe, A\BPBI R.%
		, Halekas, J\BPBI S.%
		, Mitchell, D\BPBI L.%
		, McFadden, J\BPBI P.%
		\BCBL {}\ \BBA {} Harada, Y.%
	\end{APACrefauthors}%
	\unskip\
	\newblock
	\APACrefYearMonthDay{2019}{}{}.
	\newblock
	{\BBOQ}\APACrefatitle {Mapping the lunar wake potential structure with ARTEMIS
		data} {Mapping the lunar wake potential structure with artemis data}.{\BBCQ}
	\newblock
	\APACjournalVolNumPages{Journal of Geophysical Research: Space
		Physics}{124}{5}{3360--3377}.
	\PrintBackRefs{\CurrentBib}
	
	\bibitem [\protect \citeauthoryear {%
		Zhu%
		, Byrd%
		, Lu%
		\BCBL {}\ \BBA {} Nocedal%
	}{%
		Zhu%
		\ \protect \BOthers {.}}{%
		{\protect \APACyear {1997}}%
	}]{%
		zhu1997algorithm}
	\APACinsertmetastar {%
		zhu1997algorithm}%
	\begin{APACrefauthors}%
		Zhu, C.%
		, Byrd, R\BPBI H.%
		, Lu, P.%
		\BCBL {}\ \BBA {} Nocedal, J.%
	\end{APACrefauthors}%
	\unskip\
	\newblock
	\APACrefYearMonthDay{1997}{}{}.
	\newblock
	{\BBOQ}\APACrefatitle {Algorithm 778: L-BFGS-B: Fortran subroutines for
		large-scale bound-constrained optimization} {Algorithm 778: L-bfgs-b: Fortran
		subroutines for large-scale bound-constrained optimization}.{\BBCQ}
	\newblock
	\APACjournalVolNumPages{ACM Transactions on mathematical software
		(TOMS)}{23}{4}{550--560}.
	\PrintBackRefs{\CurrentBib}
	
\end{thebibliography}

%Reference citation instructions and examples:
%
% Please use ONLY \cite and \citeA for reference citations.
% \cite for parenthetical references
% ...as shown in recent studies (Simpson et al., 2019)
% \citeA for in-text citations
% ...Simpson et al. (2019) have shown...
%
%
%...as shown by \citeA{jskilby}.
%...as shown by \citeA{lewin76}, \citeA{carson86}, \citeA{bartoldy02}, and \citeA{rinaldi03}.
%...has been shown \cite{jskilbye}.
%...has been shown \cite{lewin76,carson86,bartoldy02,rinaldi03}.
%... \cite <i.e.>[]{lewin76,carson86,bartoldy02,rinaldi03}.
%...has been shown by \cite <e.g.,>[and others]{lewin76}.
%
% apacite uses < > for prenotes and [ ] for postnotes
% DO NOT use other cite commands (e.g., \citet, \citep, \citeyear, \nocite, \citealp, etc.).
%

\end{document}